\documentclass[twocolumn]{aastex631}
\usepackage{amsmath,amssymb}
\usepackage{tablefootnote}

\newcommand{\refeq}[1]{Eq.~(\ref{eq:#1})}          
          
\newcommand{\reffig}[1]{Figure~\ref{fig:#1}}          
\newcommand{\refsec}[1]{Section~\ref{sec:#1}}

\newcommand{\reftab}[1]{Table~\ref{tab:#1}}

\newcommand{\be}{\begin{equation}}
\newcommand{\ee}{\end{equation}}
\newcommand{\ba}{\begin{eqnarray}}
\newcommand{\ea}{\end{eqnarray}}

\newcommand{\nele}{n_{\rm e}}

\newcommand{\kB}{k_{\rm B}}
\newcommand{\Te}{T_{\rm e}}
\newcommand{\Ebv}{E({\rm B}-{\rm V})}

\newcommand{\bfvs}{{\boldsymbol{v}_{\rm s}}}
\newcommand{\bfvl}{{\boldsymbol{v}_{\rm l}}}
\newcommand{\bfvo}{{\boldsymbol{v}_{\rm o}}}

\newcommand{\Ds}{{D_{\rm s}}}
\newcommand{\Dl}{{D_{\rm l}}}
\newcommand{\Dls}{{D_{\rm ls}}}

\newcommand{\rmd}{{\rm d}}

\defcitealias{Molla2012StarClusterEjecta}{MT12}
\defcitealias{Pascale2023}{P23}

\shorttitle{Godzilla of Sunburst Arc}
\shortauthors{Pascale and Dai}

\begin{document}

\title{A Young Super Star Cluster Powering a Nebula of Retained Massive Star Ejecta}

\author[0000-0002-2282-8795]{Massimo Pascale}
\affiliation{Department of Astronomy, University of California, 501 Campbell Hall \#3411, Berkeley, CA 94720, USA}

\author[0000-0003-2091-8946]{Liang Dai}
\affiliation{Department of Physics, University of California, 366 Physics North MC 7300, Berkeley, CA. 94720, USA}





\begin{abstract}

We suggest that ``Godzilla'' of the lensed Sunburst galaxy ($z=2.37$) is a young super star cluster powering a nebula of gravitationally trapped stellar ejecta. 
Employing HST photometry and spectroscopy from VLT/MUSE and VLT/X-Shooter, we infer physical and chemical properties of the cluster and nebula, finding Godzilla is young $4$—$6\,$Myr, massive $2 \times 10^{6}\,M_\odot\,(1000/\mu)$, a stellar metallicity $Z \simeq 0.25\,Z_\odot$, and has a compact FUV component $\lesssim 1\,{\rm pc}\,(1000/\mu)$, where $\mu$ is the flux magnification factor. The gas is significantly enriched with N and He, indicating stellar wind material, and has highly elevated O relative to the stellar metallicity, indicating entrainment of CCSNe ejecta. The high density $n_{\rm e} \simeq 10^{7-8}\,{\rm cm}^{-3}$ implies a highly pressurized intracluster environment. We propose the pressure results from CCSN-driven supersonic turbulence in warm, self-shielding gas, which has accumulated in the cluster center after runaway radiative cooling and is dense enough to resist removal by CCSNe. The nebula gas shows sub-solar C/O, Ne/O and Si/O, which may reflect the CCSN element yields for initial stellar masses $>40\,M_\odot$. A comparison to element yield synthesis models for young star clusters shows the gas abundances are consistent with complete retention and mixture of stellar winds and CCSNe ejecta until the inferred cluster age. The inferred O and He enhancement may have implications for the formation of multiple stellar populations in globular clusters, as stars formed from this gas would contradict the observed abundances of second-population stars.

\end{abstract}



\section{Introduction}
\label{sec:intro}

\begin{figure}[t]
    \centering
    \includegraphics[width=\columnwidth]{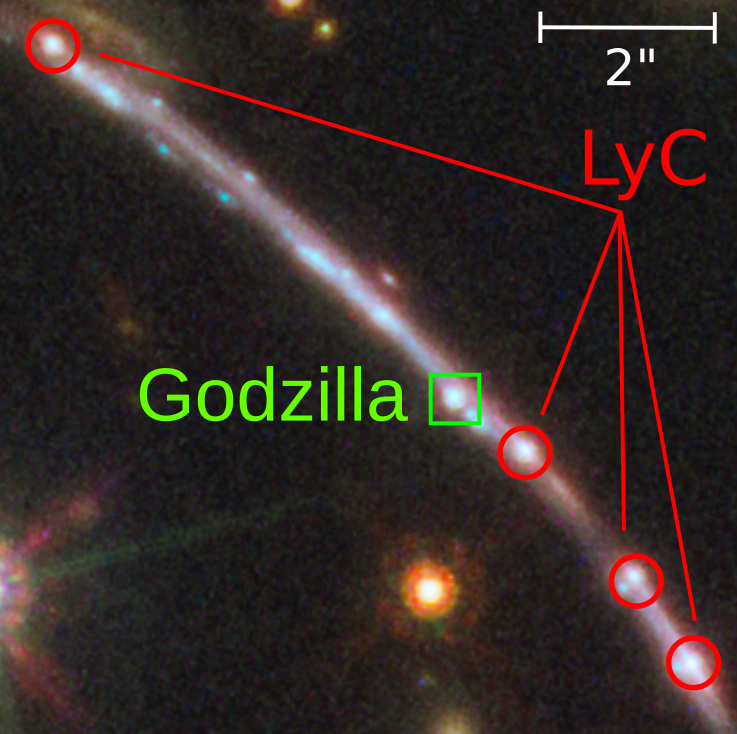}
    \caption{{\it HST} false color image of the Sunburst Arc with the lone image of Godzilla and four images of the Lyman-continuum-leaking (LyC) cluster marked. Godzilla appears similarly bright to the multiple images of the LyC cluster, so naively multiple counter-images of Godzilla would be detectable in a typical strong-lensing configuration. However, the non-detection of counter-images and time variability favors the hypothesis that Godzilla is milli-lensed and highly magnified.}
    \label{fig:color_im}
\end{figure}

The Sunburst Arc, a Cosmic Noon Lyman-$\alpha$ emitter at $z_s=2.37$~\citep{Dahle2016SunburstDiscovery} in the lensing field of the galaxy cluster PSZ1 G311.65-18.48 at $z_l=0.443$~\citep{Planck2014SZClusterCatalog}, is a spectacular gravitationally magnified galaxy~\citep{RiveraThorsen2017DirectLymanAlpha, RiveraThorsen2019Sci}. With spectroscopic confirmation, the arc hosts an apparently unresolved source bright in rest-frame FUV ($m_{\rm F814W} \approx 22$; Figure~\ref{fig:color_im}), which has been given the nickname ``Godzilla''~\citep{Diego2022godzilla}. 

Godzilla exhibits several enigmatic aspects. Much to the surprise of lens modellers, it has no confirmed counter lensed images elsewhere on the arc with comparable magnitudes, despite that multiple lens models predict so~\citep{Vanzella2020IonizeIGM, Pignataro2021SunburstLensModel, Diego2022godzilla, Sharon2022SunburstLensModel}. This implies a compact source size, with a flux enhanced by invisible milli-lenses such as sub-galactic dark matter halos~\citep{Dai2020S1226millilens, Diego2022godzilla, Diego2023mothra}.
Moreover, Godzilla shows a highly unusual nebular spectrum from rest-frame FUV to optical. \cite{Vanzella2020Tr} first inferred extremely high electron density for the ionized gas $\nele\gtrsim 10^6\,{\rm cm}^{-3}$ from ${\rm C III]}\lambda\lambda 1908,1906$ and ${\rm Si III]}\lambda\lambda 1892,1883$ line ratios, and commented on the striking weakness of hydrogen Balmer lines~\citep{Vanzella2020Tr}. \cite{Vanzella2020Tr} also detected rare Bowen fluorescence of Fe III lines pumped by Ly$\alpha$ radiation, implying an optically thick gaseous environment that efficiently traps but does not destroy Ly$\alpha$ photons. 

\cite{Diego2022godzilla} considered multiple possibilities for the underlying nature of Godzilla, which include a bright stellar transient such as an ongoing luminous blue variable (LBV) magnified by $\sim 10^4$ fold, a hyper-luminous supermassive star, or a luminous accretion disk around an intermediate mass black hole. Stringent limits on continuum photometric variability over the timescale of $\sim 2\,$yr in the source frame constrain the transient scenario. Following the submission of this paper, a preprint by \cite{Choe2024} appeared, presenting new JWST NIRCam and NIRSpec IFU data. \cite{Choe2024} further explore the LBV scenario, detecting broad components to optical emission lines which were not detected from prior ground-based data. \cite{Choe2024} conclude that Godzilla may be a binary system consisting of a post-outburst LBV with an O4-O5 supergiant companion, mirroring the $\eta$ Carinae system, however do not provide an investigation of the spectral energy distribution (SED) nor assessment of the necessary lensing magnification.

In this paper, we suggest that Godzilla is likely a compact young massive star cluster powering nebular emission, which we deem more probable as Sunburst is found to be vigorously forming stars in clusters~\citep{Vanzella2022SunburstEfficiency}. Compared to the Lyman-continuum-leaking (LyC) star cluster in the same galaxy which has been extensively studied~\citep{Chisholm2019ExtragalacticMassiveStarPopulation, Vanzella2022SunburstEfficiency, Mainali2022OutflowLyCescape, Pascale2023, Kim2023, Mestric2023VMS, RiveraThorsen2024}, Godzilla shows a similar shape of the FUV continuum as well as prominent stellar wind features tracing O stars (e.g. C IV$\lambda$1550 P Cygni profile; \reffig{MUSE_FUV_spectrum}), lending support to the star cluster interpretation. Narrow (${\rm FWHM}\lesssim 100\,{\rm km}\,{\rm s}^{-1}$) emission lines of various metal ions (O II, O III, N III, C III, Ne III, Si III) are consistent with photoionization by O stars within a star-cluster-scale gravitational potential, but ions requiring more energetic ionizing sources such as an AGN are unseen. If our model is correct, detailed study of Godzilla will greatly deepen our understanding of feedback and self-enrichment from clustered star formation in dense environments at the culmination of cosmic star formation.

In this work, we perform a joint photometric and spectroscopic analysis using archival Hubble Space Telescope (HST) imaging and ground-based Very Large Telescope (VLT) MUSE and VLT/X-shooter spectroscopy. To infer the physical parameters of the system and the chemical abundances of the nebular gas, we follow a methodology which was successfully applied to the LyC cluster of the same galaxy in \cite{Pascale2023} (hereafter \citetalias{Pascale2023}), fitting a spectral energy distribution (SED) to match broad-filter photometry and nebular emission line fluxes. Informed by this inference, we then construct a self-consistent physical model supported by the data.

Our modeling suggests that the cluster has an age $\sim 4$--$6\,$Myr and has a stellar metallicity similar to the Small Magellanic Cloud (SMC), consistent with the expectations from observed stellar wind features \citep{Chisholm2019ExtragalacticMassiveStarPopulation}. Electron density diagnostics including the ${\rm C III]}\lambda\lambda 1908,1906$ and ${\rm Si III]}\lambda\lambda 1892,1883$ line ratios and the relative strengths between {\rm [O III]}$\lambda\lambda$4959,5007 and {\rm [O III]}$\lambda\lambda$1660,1666 confirm an unusually high density $\nele \sim 10^{(7-8)}\,{\rm cm}^{-3}$, corresponding to a pressure on the order $\log P\,[{\rm K~cm}^{-3}] \sim 11$--$12$. Exciting such dense gas requires an enormous hydrogen ionizing photon flux $\log\Phi({\rm H}^0)\,[{\rm s}^{-1} {\rm ~cm}^{-2}]\sim 15$--$16$, which constrains the nebula to be smaller than $\sim 1\,$pc and likely places it within the cluster. Remarkably, we find evidence of gas-phase abundance enhancement in N, C, O, and He, such that the nebula gas exhibits solar-like O/H, high N/O similar to high-$z$ N-emitters~\citep{MarquesChavez2024}, yet sub-solar C/O, Si/O and Ne/O. We suggest this is due to self-enrichment by massive star winds and core-collapse supernova (CCSN) ejecta, and corroborate this hypothesis through a comparison with synthetic light element yield models.

\begin{figure*}[t]
    \centering
    \includegraphics[width=\textwidth]{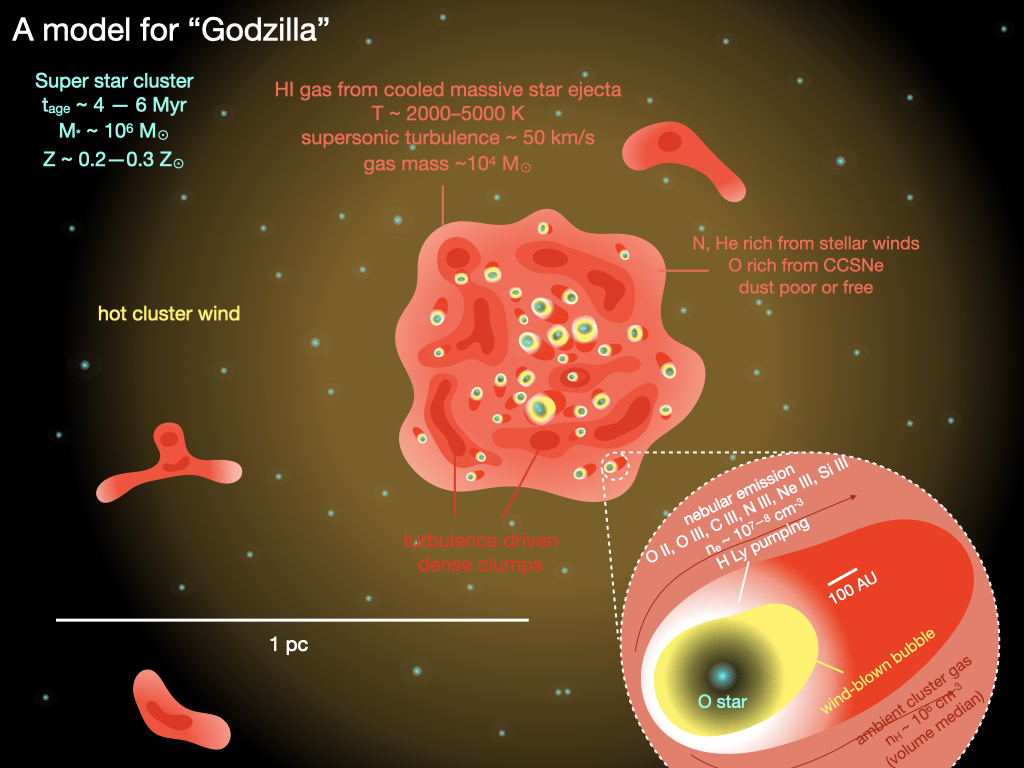}
    \caption{Illustration of our favored physical model for Godzilla and its nebula. Due to highly efficient radiative cooling, massive star ejecta from winds and CCSNe has accumulated at the center of a massive, compact young star cluster, forming self-shielded neutral warm gas in a state of supersonic turbulence driven by energy injection from CCSNe. Situated in an extreme radiation environment, this gravitationally retained gas likely has little to no dust. As a result, it is optically thick to ionizing photons but is transparent to UV and optical photons. O stars embedded inside it drive individual wind-blown hot bubbles in which the observed nebula emission lines are excited. In the outskirts of the cluster, stars likely drive a hot, low density cluster wind which may accelerate warm cloudlets resulting in a secondary broad nebular emission component.}
    \label{fig:cartoon}
\end{figure*}

These remarkable properties may be a natural consequence of young star cluster evolution in the regime of high mass and high compactness. We develop a physical picture where the ejecta of winds and CCSNe stays inside the cluster through rapid radiative cooling \citep{Wunsch2011ClusterWindWithCooling, Wunsch2017RapidCoolingRHDsimulation}, but is dynamically supported by supersonic turbulence driven by CCSNe. This physical picture, depicted in \reffig{cartoon}, congruently interprets the unusual dataset; it can simultaneously explain the observed high electron density, high pressure, ionization parameter, and chemical abundances indicative of retention of stellar ejecta, and may provide the required gas geometry for Ly$\alpha$-pumped Fe fluorescence.

The remainder of this paper will be organized as the following. In \refsec{data}, we summarize the public imaging and spectroscopic datasets used in this work. In \refsec{phot} and \refsec{sed}, we discuss how we will model broad-band photometry and emission lines for the star cluster and its nebula by applying BPASS stellar population synthesis models and \texttt{Cloudy} photoionization calculations. In particular, we will discuss how we extend our \texttt{Cloudy} calculations to allow for non-standard gas-phase element abundances. In \refsec{results}, we present results for joint photometric and emission line fitting for two models and infer key physical and chemical properties of the star cluster and its ionized gas. In \refsec{discuss}, we discuss the implications of our fitting results for the dynamics, chemical composition, and astrophysical origin of Godzilla's nebula. In Section \ref{sec:mag}, we explore the possible magnification of Godzilla, setting upper limits on the basis of flux variability, and evaluating the probability of Godzilla being a single highly magnified, hyperluminous star. Concluding remarks will be provided in \refsec{concl}.

\section{Data}
\label{sec:data}

In this section, we summarize the archival data used in this work: HST imagining data, VLT/MUSE IFU spectroscopy, and VLT/X-shooter slit spectroscopy.

\subsection{{\textit HST} Imaging}

Observations for PSZ1 G311.65-18.48 were obtained across multiple programs between 2018 and 2020 (P.I. Dahle Program ID 15101, P.I. Gladders Program ID 15949, P.I. Bayliss Program ID 15377). In this work, we make use of observations publicly available on the MAST archive for 15 filters: WFC3/UVIS F390W, WFC3/UVIS F410M, WFC3/UVIS F555W, WFC3/UVIS F606W, ACS/WFC F814W, WFC3/UVIS F098M, WFC3/IR F105W, WFC3/IR F125W, WFC3/IR F126N, WFC3/IR F128N, WFC3/IR F140W, WFC3/IR F153M, WFC3/IR F160W, WFC3/IR F164N, WFC3/F167N. All UVIS+WFC images were aligned to a common pixel scale of 30 mas/pixel, while all IR images were aligned to a pixel scale of 60 mas/pixel using the astropy \texttt{reproject} package in combination with \texttt{astroalign} \citep{Beroiz2020}.

\subsection{VLT/MUSE IFU Spectroscopy}

We use reduced MUSE IFU science datacubes publicly accessible from the ESO archive\footnote{\url{http://archive.eso.org/scienceportal/home}}. Two datacubes are used, which cover the observed wavelength range $475$--$935\,$nm at $R\approx 2600$--$3000$ and were calibrated and reduced with the standard reduction pipeline: one obtained on May 13 and Aug 24 in 2016 (Program 107.22SK; PI: E. Vanzella; total exposure $4449\,$s), and another on Aug 10, 2021 (Program 297.A-5012; PI: N. Aghanim; total exposure $2808\,$s). The MUSE IFU observations were carried out in the Wide Field Mode under a typical seeing condition FWHM $= 0.5$--$0.6\arcsec$, achieving a spatial resolution $1.4$--$2\arcsec$ on the sky. Through a comparison between the two datacubes, we do not find any evidence for spectral variability, so we properly combine them to achieve the best signal-to-noise ratio for emission lines.

\subsection{VLT/X-shooter Slit Spectroscopy}

X-shooter slit spectroscopy data reduced with the standard reduction pipeline are accessed from the ESO archive. Data were collected from Program 0103.A-0688 (PI: E. Vanzella). Exposures were obtained under typical seeing conditions ${\rm FWHM}\sim 0.5\mbox{--}0.6\arcsec$. Slit spectra covering observed 994--2479 nm (R$\approx$8000) are used for extracting fluxes for rest-frame NUV and optical emission lines, while additional spectra covering observed 534--1020 nm (R$\approx$11000) are used for cross-checking strong FUV emission lines. For accessing spatially resolved spectral information, we carry out flux calibration of the 2D spectra by comparing the flux-calibrated and flux-uncalibrated 1D spectra, and then extract fluxes at the location of Godzilla.

For both MUSE and X-Shooter, the line flux uncertainty is calculated by line injection in the continuum neighboring each line followed by Gaussian fitting. For most optical lines in X-Shooter, candidate signal is visible at the location of Godzilla along the slit, but blending with strong line emissions from the neighboring Image 8 of the LyC cluster is too severe. Therefore, detection is conservatively quoted as an upper limit. Further details on extraction of the emission line fluxes, aperture correction and uncertainty estimation for the VLT/MUSE and VLT/X-Shooter spectra can be found in Section 2 of \citetalias{Pascale2023}.


\begin{table}
	\centering
	\caption{Inferred isotropic luminosities for nebular emission lines detected from VLT/MUSE IFU data for Godzilla. The isotropic luminosity is calculated for a luminosity distance of $d=19.566\,$Gpc (corresponding to the source redshift $z_s=2.369$) and a fiducial achromatic magnification factor $\mu=300$, and is uncorrected for dust reddening.}
	\label{tab:MUSE_lines}
	\begin{tabular}{rc}
		\hline
		\hline
		Emission line & Luminosity $L\,[10^{38}\,{\rm erg}\,{\rm s}^{-1}]$ \\
		\hline
		{\rm O I}$\lambda$1641 & $26\pm 4$ \\
        {\rm [O III]}$\lambda$1660 & $21\pm 7$ \\
        {\rm [O III]}$\lambda$1666 & $55\pm 6$ \\
        {\rm N III]}$\lambda$1750 & $73\pm 12$  \\
        {\rm N III]}$\lambda$1752 & $29\pm 5$  \\
        {\rm [Si III]}$\lambda$1883 & $\lesssim 8$ \\
        {\rm Si III]}$\lambda$1892 &  $54\pm 4$\\
        {\rm [C III]}$\lambda$1906 & $\lesssim 10$\\
        {\rm C III]}$\lambda$1908 & $123\pm 8$\\
        {\rm [O II]}$\lambda$2471 & $14\pm 4$ \\
		\hline
        \hline
	\end{tabular}
\end{table}

\begin{table}
	\centering
	\caption{Inferred isotropic luminosities of nebular emission lines detected from VLT/X-shooter for Godzilla. The same luminosity distance and fiducial magnification factor $\mu=300$ are used as in \reftab{MUSE_lines}, while dust reddening is not corrected for. We have applied a wavelength independent PSF correction factor $=1.2$ using a theoretical Moffat PSF model with $\beta=4.765$.}
	\label{tab:Xshooter_lines}
	\begin{tabular}{rc}
		\hline
		\hline
		Emission line & Luminosity $L\,[10^{38}\,{\rm erg}\,{\rm s}^{-1}]$ \\
		\hline
		Ly$\alpha$ & $\lesssim 240$ \\
        {\rm [O II]}$\lambda$3726 & $\lesssim 42$ \\
        {\rm [O II]}$\lambda$3729 & $\lesssim 24$ \\
        {\rm [Ne III]}$\lambda$3869 & $37\pm 7$ \\
        {\rm [Ne III]}$\lambda$3967 & $\lesssim 12$ \\
        H$\beta$ &  $\lesssim 61$\\
        {\rm [O III]}$\lambda$4959 & $\lesssim 61$\\
        {\rm [O III]}$\lambda$5007 & $\lesssim 180$\\
        H$\alpha$ & $\lesssim 185$ \\
        {\rm He I}$\lambda$5876 & $36\pm 18$ \\
		\hline
        \hline
	\end{tabular}
\end{table}

\begin{figure*}[ht]
    \centering
    \includegraphics[scale=0.43]{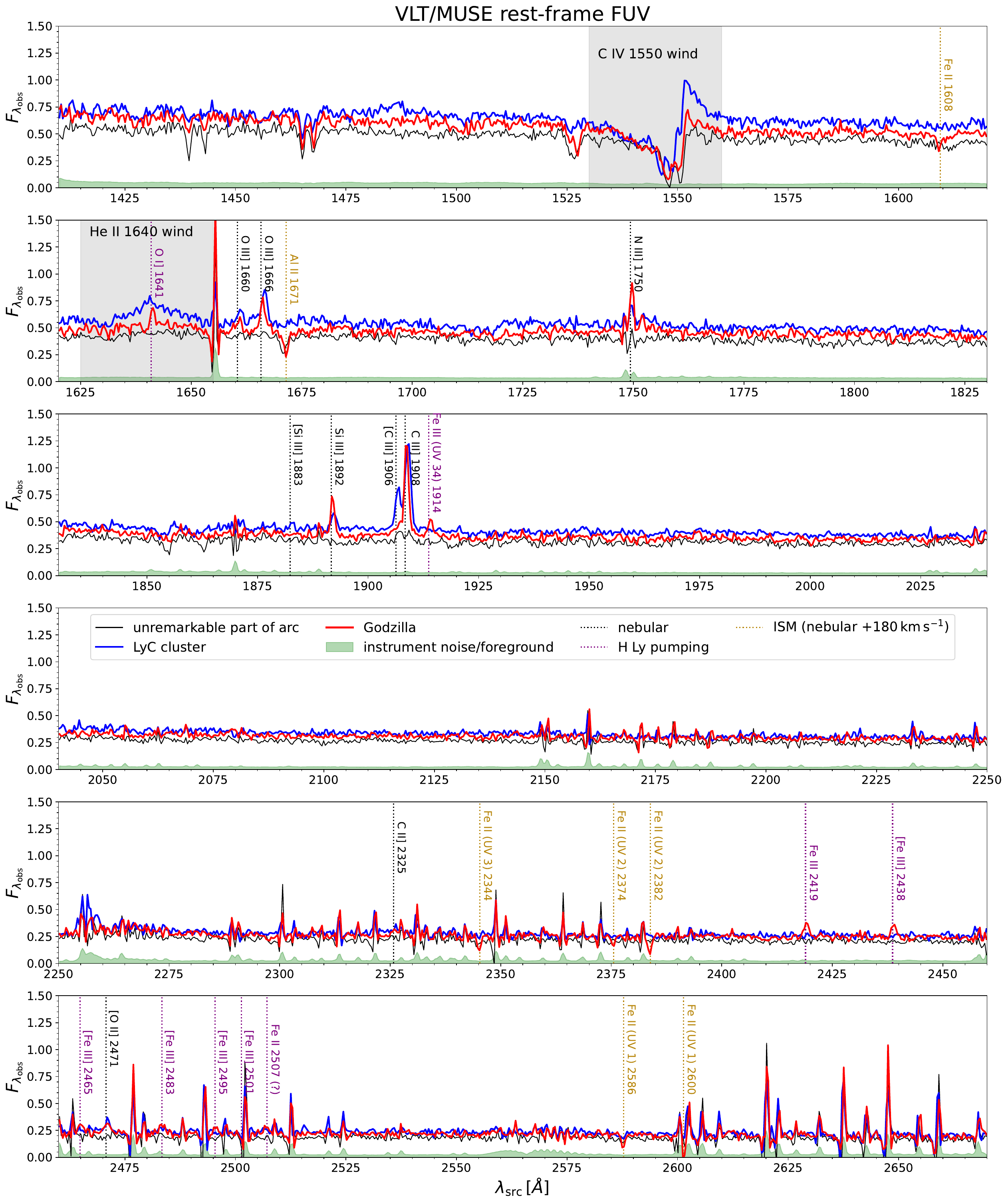}
    \caption{MUSE IFU spectrum in the rest-frame wavelength range $1410$--$2670\,\AA$ (observed flux density $F_{\lambda_{\rm obs}}$ in units of $10^{-17}\,{\rm erg}\,{\rm s}^{-1}\,{\rm cm}^{-2}\,{\AA}^{-1}$). We compare specific fluxes extracted from three circular apertures of radius $0.5''$, centered on {\it Godzilla} (red), the LyC leaking star cluster~\citep{Vanzella2020IonizeIGM} (blue), as well as another unremarkable region on the Sunburst Arc (black), respectively. We mark narrow nebular emission lines (black dotted), ISM absorption lines (dotted yellow), and additional unusual emission lines related to H Lyman pumping (dotted magenta).}
    \label{fig:MUSE_FUV_spectrum}
\end{figure*}

\section{Photometry}
\label{sec:phot}

Following the methods of \citetalias{Pascale2023}, we measure the PSF of each HST image using the \texttt{photutils} package by stacking isolated, unsaturated stars in the field. An initial catalog of stars is generated using \texttt{DAOStarFinder} following \cite{Stetson1987}, and then each star is vetted by eye to be isolated and unsaturated, resulting in a final catalog of dozens of stars. The final star list is extracted into centered cutouts which are normalized and median-subtracted to mitigate sky background, and the cutouts are stacked into an oversampled PSF using the \texttt{EPSFBuilder} function.

\begin{figure}[t]
    \centering
    \includegraphics[width=8.5cm,clip]{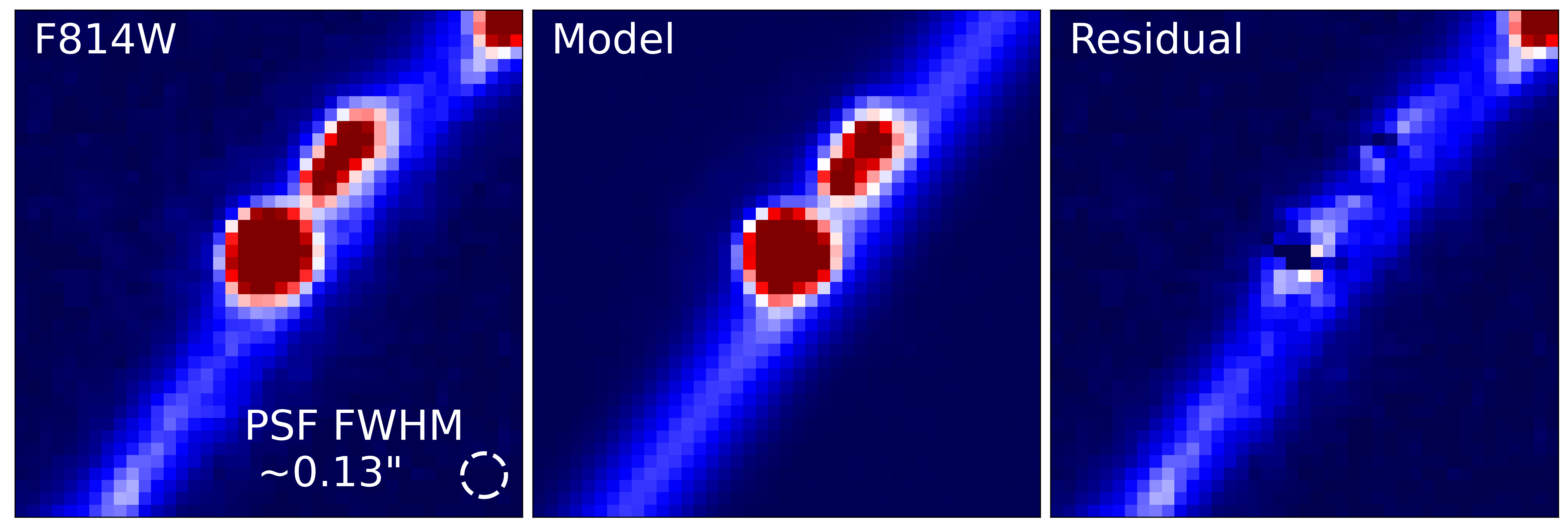}
     \caption{Example of surface brightness modeling of Godzilla and the host arc in F814W. Godzilla is modeled as a S\'{e}rsic profile convolved with the PSF, and its neighboring knots are modeled as simple PSFs. The underlying arc background is approximated as a line convolved with a 2D-Gaussian, for which the line slope and intercept, as well as the Gaussian width are free parameters. We find that, even in the highest resolution filters, the best fit S\'{e}rsic for Godzilla has effective radius $r_{\rm eff} <1 {\rm ~pixel}$, consistent with an unresolved FUV source. For a fiducial tangential magnification value $\mu_t = 500$, this implies a physical size for the OB star component $R_{\rm FUV} < 1$~pc given the measured F814W PSF FWHM of $0.13\arcsec$.}
    \label{fig:astrometric}
\end{figure}

To measure fluxes for Godzilla, we model the surface brightness profile as a S\'{e}rsic profile convolved with the PSF \citep{Sersic1968}. There are two primary sources of contamination which complicate fitting: neighboring star-formation knots on the arc and the diffuse arc background itself. Each of the neighboring knots are observed to be consistent with PSF profiles and are simultaneously fit in PSF photometry. The arc background is modeled as a thick line of smoothly-varying surface brightness, with free parameters for slope, intercept, and normalization. This simple parametrization is found to well approximate the arc background locally (within $\sim 1-2^{\prime \prime}$), and is fit simultaneously to Godzilla and neighboring knots. As an example, the best fit model and residual for the F814W filter is shown in Figure~\ref{fig:astrometric}.

Across all HST images, the best-fit S\'{e}rsic profiles for Godzilla consistently have $r_{\rm eff} \lesssim 1$~pixel, implying that Godzilla is unresolved and consistent with the PSF (we measure FWHM$\sim 0.13\arcsec$ in F814W, consistent with \cite{Vanzella2020IonizeIGM}). This constrains the spatial extent $R_{\rm FUV}\lesssim 1\,{\rm pc}\,(500/\mu_t)$ of the OB star population, for a given tangential magnification $\mu_t$. The tangential magnification is used in place of the total magnification as the source is preferentially stretched in the tangential direction. Indeed lens modeling studies find the radial magnification along the arc is likely modest $\mu_r \sim 2$, where $\mu = \mu_t\,\mu_r$ \citep{Diego2022godzilla}. The statistical errors from fitting are likely subdominant to systematics involving parametrization of the various components -- to account for this we set a conservative photometric error of $0.05\,$mag for the WFC3/UVIS and ACS/WFC filters and $0.1\,$mag for the WFC3/IR filters. These were assigned to be slightly larger than the systematics found from injection recovery tests of the PSF on blank (no bright point sources) regions of the arc, as these test were not robust to systematics involving the PSF itself.

\begin{table}
	\centering
	\caption{HST photometry for Godzilla. Lensing magnification is uncorrected for.}
	\label{tab:photometry}
	\begin{tabular}{rcl}
		\hline
		\hline
		Filter & AB magnitude & Spectral features \\
		\hline
            F275W$^{a}$ & $\gtrsim 28.6$ & Rest-frame LyC; \\
		F390W$^{a}$ & $22.41 \pm 0.10$ & Ly$\alpha$ \& damping wing \\
		F410M$^{a}$ & $22.88 \pm 0.10$& Ly$\alpha$ \& damping wing \\
            \hline
		F555W & $22.00 \pm 0.05$ & -- \\
		F606W & $22.00 \pm 0.05$ & --\\
		F814W & $22.00 \pm 0.05$ &  --\\
		F098M & $22.18 \pm 0.10$ & --\\
		F105W & $22.16 \pm 0.10$ & --\\
            F125W & $22.18 \pm 0.10$ & -- \\
		F126N & $22.16 \pm 0.10$ &[O II]$\lambda\lambda$3727,3729 \\
             F128N & $22.24 \pm 0.15$ & --\\
		F140W & $22.18 \pm 0.10$ & --\\
		F153M & $22.28 \pm 0.10$ &--\\
		F160W & $22.17 \pm 0.10$ & H$\beta$, [O III]$\lambda\lambda$4959,5007 \\
		F164N$^{a}$  & $22.17 \pm 0.10$ & H$\beta$ \\
            F167N & $22.05 \pm 0.10$ & [O III]$\lambda$4959 \\ 
		\hline
        \hline
	\end{tabular}
\tablenotetext{a}{Excluded from analysis.}
\end{table}
\section{Spectral Modeling}
\label{sec:sed}

To unveil the nature of Godzilla and the origin of its nebular emission, we study several physical models describing a young mass star cluster associated with a photoionization nebula. We will adopt an iterative strategy. We shall first study simple models; if significant discrepancy with data is identified, we will be guided to consider models describing more unusual but physically motivated situations.

To that end, we will perform joint photometric-spectroscopic fitting using photometry from 12 HST filters (\reftab{photometry}) and 18 emission line flux detections or upper limits from VLT/MUSE and VLT/X-shooter (\reftab{MUSE_lines} and \reftab{Xshooter_lines}). We shall include only 12 of the 15 available HST filters in fitting. Photometry in F390W and F410M are sensitive to Ly$\alpha$ line transfer  and damping wing effects, which we do not attempt to carefully model. We also exclude F164N; while this filter constrains H$\beta$, we are aware of a flux calibration issue in this filter~\citep{Kim2023}.

While several filters cover emission lines including ${\rm [O II]}\lambda\lambda3726,3729$ (F126N), H$\beta$ (F160W, F164N), and ${\rm [O III]}\lambda\lambda4959,5007$ (F160W), they appear weak and only flux upper limits are reliable. We note that HST photometry is significantly more sensitive and hence may constrain these emission lines better than X-shooter data.

Emission line measurements prove highly valuable in probing the properties of Godzilla's nebula gas. Even many upper limits are surprisingly informative. The C III]$\lambda\lambda$1908,1906 and Si III]$\lambda\lambda$1892,1883 line ratios probe the electron density in a high ionization environment. Non-detection of the forbidden component~\citep{Vanzella2020Tr} indicates that Godzilla has a nebula much denser than what are often considered dense high-$z$ HII regions with $3 \lesssim \log \nele\,[{\rm cm}^{-3}] \lesssim 6$ \citep{Keenan1992CIIISiIII, Jaskot2016CIII}. Ratios between the O II and O III lines ${\rm [O II]} \lambda 2471 / {\rm [O III]}\lambda\lambda 1660,1666$ \citep{Kewley2019ARAAreview} are sensitive to ionization degree and electron temperature \citep{KewleyDopita2002LineDiagnostics, KobulnickyKewley2004, Dors2011}. For Godzilla, the ${\rm [O III]}\lambda\lambda 1660,1666 / {\rm [O III]} \lambda\lambda 4959,5007$ line ratios also constrain the electron density, as these trace the same ion but the rest-frame optical doublet has a much lower critical density at $6.4\times 10^5\,{\rm cm}^{-3}$~\citep{Draine2011ISMtext}.

\subsection{Base Model}

We first study a model describing photoionized gas excited by the entire star cluster. This has been used to model the spectrum of the Lyman-continuum-leaking cluster of the Sunburst Arc in \citetalias{Pascale2023}.
The model star cluster SED is taken from BPASS (v2.0) with binary stellar evolution accounted for~\citep{Stanway2016BPASS}. For a compact, massive star cluster, we consider an instantaneous burst of star formation with an initial mass function (IMF) $\xi(m) \propto m^{-1.3}$ for $m<0.5\,M_\odot$ and $\xi(m) \propto m^{-2.35}$ for $m>0.5\,M_\odot$ in the initial stellar mass range $0.1 < m/M_\odot <300$. The cluster is characterized by the age $t_{\rm age}$ and stellar metallicity $Z$ as free parameters.

Following the method of \citetalias{Pascale2023}, nebular continuum, emission lines, and attenuation of incident star light are all modeled with the \texttt{Cloudy} code \citep[v17][]{Ferland2017c17} assuming a plane-parallel, stratified structure locally. The nebula gas is assumed to be isobaric, such that it is characterized by the typical dimensionless ionization parameter $U$ and the constant pressure $P$. In Bayesian inference, we shall use log-flat priors for these, in the range $-3\leq\log U \leq -1$ and $6 \leq \log P\,[{\rm K~cm}^{-3}] \leq 13$. 

When running \texttt{Cloudy} calculations for the Base Model, we identify the stellar metallicity $Z$ with the gas-phase metallicity and use a rescaled solar abundance pattern (see Table 7.1 or the \texttt{Cloudy} manual \texttt{HAZY} 1), with non-rescaled-solar prescriptions for He \citep[following ][]{RussellDopita1992abunMagellanic}, C, and N \citep[following][]{Dopita2006SEDstarburst}. The abundances of N, C, Ne, and Si are allowed to freely vary only when we fit the emission line data; we do this by rescaling the fluxes of emission lines corresponding to these elements. This is a computationally efficient and justifiable approximation provided that varying the abundance of these elements does not significantly change line cooling. 

We also introduce two geometric parameters: the parameter $x$ ($0 \leqslant x \leqslant 1$) is the fraction of ionizing radiation processed by the nebular gas. A second parameter, $y$, describes the fraction of nebular gas toward which our sight-line is intervened by HI gas ($y=0$) versus the fraction toward which the sight-line is not ($y=1$). We shall refer to the former case as ``HI-obscured'' (see Figure 2 of \citetalias{Pascale2023}). For example, the latter case can correspond to directly seeing the irradiated side of the HII zone on the surface of a self-shielding cloud, without any intervening gas. The former case can correspond to viewing from the back side of the self-shielding cloud, so that photons reaching the observer must pass through a substantial HI column of the cloud. Without internal dust attenuation, the parameter $y$ does not affect (semi-)forbidden lines, but has an impact on lines with a significant optical depth effect or with a large diffusive contribution in the HI layer.

\citetalias{Pascale2023} introduced a third geometric parameter, $z$, which measures the fraction of star light along the line-of-sight that is obscured by photoionized gas. Running \texttt{Cloudy} models without internal dust, we do not find it strongly constrained by data and hence set it to zero obscuration. Finally, we also incorporate external dust reddening, fixing Milky Way extinction following the galactic dust maps of \cite{Schlegel1998}, and fitting for host galaxy ISM dust reddening following the law of \cite{Reddy2015DustRedden}, where the normalization \Ebv \,is left free. In total, the Base Model has 10 free parameters.


\subsection{Chemically Anomalous Model}
\label{sec:chemanomaly}

In Godzilla, some unusual emission line ratios hint at the possibility of non-standard gas-phase element abundances. Apart from a clear detection of N enhancement (like what we found for the LyC cluster in \citetalias{Pascale2023}), we see evidence for He and O enhancement, as well as unusual C/O and Ne/O ratios. The surprising weakness of H Balmer lines relative to metal forbidden lines~\citep{Vanzella2020Tr} may also be related to an unusual abundance pattern. 

Chemical self-enrichment is physically plausible for newborn massive star clusters if massive star ejecta can be retained in the cluster vicinity or interior \citep{deMink2009BinaryAbundanceAnomaly, Martell2013, LochhaasThompson2017Cooling}. Massive star winds or envelope stripping may cause self-enrichment of N and He, while CCSNe produce $\alpha$-elements including O, Ne and Si.
We are therefore motivated to consider more sophisticated non-rescaled-solar abundance patterns.
In particular, the gas-phase oxygen abundance may be enhanced compared to the interior of stars.
Unlike N, C, Si, and Ne, oxygen ions play a significant role in line cooling, and hence changes in oxygen abundance cannot be simply accounted for by rescaling O line fluxes in post-processing. 

In the Chemically Anomalous Model, the stellar SED is the same as in the Base Model.  We fix the stellar and gas-phase (before self-enrichment) metallicities to $0.25\,Z_\odot$ following the best-fit Base Model. This gas-phase metallicity is consistent with what we found for the LyC cluster of Sunburst in \citetalias{Pascale2023}, and within the inference uncertainty agrees with the gas metallicity determined by \cite{RiveraThorsen2024}. We then compute a grid of \texttt{Cloudy} models for which O/H varies from one-fourth to twice the solar value (assuming $12+\log({\rm O/H})_\odot = 8.69$)~\citep{Asplund2021SolarAbund2020}. 
Similarly, He abundance variation should be consistently treated with \texttt{Cloudy}. The detection of the recombination line {\rm He I}$\lambda$5876 probes He/H, which may be enhanced by evolved star winds. We create a He/H grid to allow enhancement up to five times greater than the prescription of \cite{RussellDopita1992abunMagellanic} at one-fourth solar metallicity. In this model, the stellar metallicity is no longer a free parameter in fitting, while gas-phase O and He abundances are now left free to vary, increasing the total number of free parameters to 11.


\section{Results}
\label{sec:results}

We fit the two models to data from 12 filters of HST photometry (Fig.~\ref{fig:sed}) and 18 emission line fluxes and upper limits (Fig.~\ref{fig:lineflux}). We use the sampler \texttt{PyMultinest} \citep{Feroz2009,Buchner2014} with 1,000 live points running for 10,000 steps to explore the model parameter space and derive posterior samples. Below we discuss the inferred physical parameters and the fitting performance for each model. We assign uncertainties in both filter photometry and emission line fluxes/upper limits conservatively, and hence the reduced chi-squared $\chi^2_\nu < 1$ obtained in our fitting is likely indicative of overestimated errorbars rather than overfitting.

\subsection{Base Model}

The Base Model well-reproduces all observables with a reduced chi-squared $\chi^2_\nu = 0.52$. However, constraints on the stellar properties are not extremely stringent. Using the BPASS instantaneous burst model, photometry of the stellar and nebular continuum indicates a high stellar mass $\log \mu M_{\star}/M_{\odot} = 9.29^{+0.11}_{-0.16}$, or $M_{\star}/M_\odot = 2 \times 10^6\,(1000/\mu)$. The magnification of Godzilla is not known as multiple lens models make a range of predictions \citep{Pignataro2021SunburstLensModel, Diego2022godzilla, Sharon2022SunburstLensModel}. The surprising absence of bright counter images imply that Godzilla's magnification factor may have been strongly boosted by milli-lensing~\citep{Diego2022godzilla}. Hereafter, we shall use a high fiducial value $\mu=1000$ and denote $\mu_{1000}=\mu/1000$ (see \refsec{mag} for more discussions of the magnification factor of Godzilla.).

The favored cluster age is in the range $t_{\rm age}=3$--$6\,$Myr. Godzilla thus appears more evolved than the LyC cluster.
As seen in Fig.~\ref{fig:MUSE_FUV_spectrum}, the emission component of the {\rm C IV}$\lambda$1550 P Cygni wind feature, whose strength is sensitive to age but not stellar metallicity \citep{Leitherer2014, Chisholm2019ExtragalacticMassiveStarPopulation}, has a significantly lower equivalent width than for the LyC cluster in the same galaxy \citep[$t_{\rm age}\sim 3\,$Myr;][]{Chisholm2019ExtragalacticMassiveStarPopulation, Pascale2023}. The higher cluster age is further corroborated by the absence of broad {\rm He II}$\lambda$1640 emission, a unique signature of Very Massive Stars ($>100\,M_\odot$) seen in the LyC cluster that would likely have died by $t_{\rm age} = 3-4\,$Myr~\citep{Mestric2023VMS}.

The stellar and gas metallicity is constrained to be $\log Z = -2.29^{+0.16}_{-0.12}$, or 20\%-40\% of the solar value. The broad posterior distribution for $Z$ is likely due to non-detections of H$\alpha$, H$\beta$, and the {\rm [O III]}$\lambda\lambda$4959,5007 doublet. Reassuringly, the median of the posterior matches the value inferred for the LyC cluster in \citetalias{Pascale2023} and \cite{RiveraThorsen2024}, as the two objects reside in the same galaxy.
For a further piece of evidence, we use the depth of the absorption component of the {\rm C IV}$\lambda$1550 P Cygni wind feature, which is primarily set by the stellar metallicity \citep{Chisholm2019ExtragalacticMassiveStarPopulation}. The depths are observed to be consistent in both Godzilla and the LyC cluster (Fig.~\ref{fig:MUSE_FUV_spectrum}).
Similar to \cite{Kim2023}, we find a minimal amount of external dust reddening ${\rm E(B-V)}=0.02^{+0.02}_{-0.02}$ based on the continuum shape.

Remarkably, the ionized gas of Godzilla has an extremely high pressure $\log P = 11.56^{+0.14}_{-0.12}\,[{\rm K~cm}^{-3}]$ and a low ionization parameter $\log U = -2.61^{+0.13}_{-0.11}$, corresponding to a characteristic incident ionizing flux $\log\Phi ({\rm H}^0) = 15.22^{+0.16}_{-0.15}\,[{\rm s}^{-1}{\rm cm}^{-2}]$ and a high electron density $\log \nele={7.12}_{-0.15}^{+0.30}\,[{\rm cm}^{-3}]$.

The Base Model allows the abundances of C, N, Ne and Si to vary freely in emission line fitting. We find all four to show unusual abundances relative to oxygen, with lower C/O, Ne/O and Si/O but significantly higher N/O compared to typical ISM values found in galaxies at sub-solar metallicity. The elevated N/O, $\log({\rm N/O}) = -0.30^{+0.07}_{-0.07}$, is reminiscent of the LyC cluster, a factor of $\sim 13$ greater than typical HII region abundances \citep{Pilyugin2012HIIRegions}. Thus, Godzilla appears to be another example of nitrogen emitters which have received heated discussion in the literature~\citep{MarquesChavez2024}. 

While the measured $\log{{\rm C/O}} = -0.79^{+0.05}_{-0.05}$ is low compared to the solar value of $-0.30$~\citep{Asplund2021SolarAbund2020}, we note that it is consistent with the $\sim -0.7$ ISM value observed in $z\sim2$--$3$ galaxies at a similar metallicity~\citep{Berg2019MetalPoorCNOEvolution} (see \S~\ref{sec:carbon}). The gas-phase Si/O appears low, $\log({\rm Si/O})=-1.84^{+0.06}_{-0.05}$, a factor of $4$--$5$ lower than the solar value $-1.15$~\citep{Asplund2021SolarAbund2020}. While grain depletion of Si may explain this, it may also point towards the intriguing possibility of self-enrichment of oxygen by CCSNe that go off in the star cluster. This, taken with the low C/O and the anomalously low measured $\log({\rm Ne/O}) = -1.20^{+0.11}_{-0.12}$ (factor of $\sim 5$ deficient relative to the solar value) motivates us to consider more sophisticated gas abundance models in which the gas-phase oxygen abundance differs from the stellar value.

The parameter $y$ is tightly constrained to be very close to zero, $y<0.03$, which implies that nearly all the nebular emission that we see is HI-obscured, mostly likely by the HI gas right behind the edge of the HII zone in an ionization-bounded geometry. The strong preference for this in data is driven by the need to significantly decrease the H Balmer lines. In \refsec{yeffect}, we discuss how H Balmer lines might be reduced by a dust-free HI zone.

\subsection{Chemically Anomalous Model}

To explore self-enrichment of gas-phase abundances, we fix the stellar metallicity to be $Z=0.25\,Z_\odot$ in the Chemically Anomalous Model, which corresponds approximately to the median of the posterior distribution of the Base Model as well as the $Z$ value inferred for the LyC cluster in \citetalias{Pascale2023}. In constructing \texttt{Cloudy} model grids, we scan a range of O/H and He/H values.
The goodness of fit is notably improved over the Base Model, achieving a lower $\chi^2_\nu = 0.39$ despite only one more free parameter. For this reason, we think that the best-fit Chemically Anomalous Model reflects reality more than the best-fit Base Model.

For most parameters, we find posterior distributions consistent with what we derive for the Base Model (Table~\ref{tab:cloudtable}). There are three notable exceptions: (1) a lower value for the effective area covering factor $x \simeq 0.2$ is now preferred; (2) the gas pressure is $\sim 0.3\,$dex higher at $\log{\rm P} \sim 11.9$; (3) the posterior distribution of the ionization parameter keeps a peak at low values $\log U \simeq -2.6$ but develops a long tail toward higher values $\log U \simeq -(1$--$2)$.

Changes in the pressure and the effective covering factor result from a preference for enhanced O/H. The strong UV {\rm [O III]}$\lambda\lambda$1660,1666 are detected, but there are interesting upper limits on the optical {\rm [O III]}$\lambda\lambda$4959,5007 and H Balmer lines. {\rm [O III]}$\lambda\lambda$4959,5007 are collisionally suppressed above $\nele = 6.4 \times 10^5\,{\rm cm}^{-3}$~\citep{Draine2011ISMtext}, such that increased $P$ will reduce the optical doublet relative to the UV doublet. By contrast, enhancing the O abundance dramatically enhances the optical [O III] doublet more than the UV [O III] doublet, likely owing to increased line cooling in a metal-rich gas preferentially suppressing the UV forbidden lines. These effects manifest as a positive correlation between $P$ and O/H, and a negative correlation between $x$ and O/H (Fig.~\ref{fig:cloudcorner}).
Since the oxygen lines increase with O/H, a solution with increased O/H is preferred to explain strong [O III] and [O II] without H Balmer detections. The covering factor must decrease to avoid overproducing oxygen lines. Because this enhances {\rm [O III]}$\lambda\lambda$4959,5007 more than {\rm [O III]}$\lambda\lambda$1660,1666, $P$ and hence $\nele$ must be increased to suppress the former.

The high $\log U$ tail appears in the posterior distribution due to a preference for enhanced He/H.
At high $\log U$, emission lines from higher excitation ions (O III, N III, Ne III) are enhanced relative to lines of lower excitation ions (O II, Si III, Fe III). Increased He abundance, which consumes more He I ionizing photons ($h\nu > 24.6\,$eV), has the opposite effect that a He I zone forms behind the He II zone where the low excitation ions can be abundant. 
Hence, there is a positive correlation between He/H and $\log U$ (Fig.~\ref{fig:cloudcorner}), as the two approximately compensate for one another with the exception of He emission lines.

As we fix the stellar metallicity $Z=0.25\,Z_\odot$ in this model, fitting results indicate that the gas-phase oxygen is enhanced by $\sim 2$--$6$ fold, producing an O/H at roughly the solar value. C, N, Si and Ne relative to O are found to be similarly abundant as in the Base Model, but have increased abundances relative to H. This implies that the inferred low C/O, Ne/O and Si/O values may be a symptom of O enrichment rather than a deficiency in C, Ne and Si. We find that He is enhanced relative to H by $\sim 2$--$4$ fold compared to the prescription of \cite{RussellDopita1992abunMagellanic}.

This model again requires $y$ to be closer to zero than to unity, but with a broader posterior distribution, $y < 0.2$, enabled by the enhanced O/H (see correlation in Figure~\ref{fig:cloudcorner}). This implies that the nebular emission should be mostly HI-obscured, but up to $\sim 20\%$ of it may still come directly from unobscured irradiated cloud surfaces. As we shall discuss later, this preference in $y$ has important implications for the probable nebula geometry (\refsec{geometry}).

Since the Chemically Anomalous Model is consistent with a wide range of $\log U$ values, we further test a model involving two ionized gas components, one with low $\log U < -1.5$ as found in the Base Model, and another with high $\log U > -1.5$.
The high ionization parameter $\log U > -1.5$ can be explained by photoionization nebulae in wind-blown bubbles surrounding individual main sequence O stars at $\sim 4\,$Myr, if the cluster interior is filled with dense self-shielding gas which could have condensed from wind and CCSN ejecta (see \reffig{cartoon} and \refsec{core}).
Assuming that all other physical parameters are the same for the two components, we find that significant processing of the total cluster ionizing output ($\gtrsim 10\%$) by a second high $\log U$ nebula is neither favored nor ruled out by data, with the other inferred physical parameters unaltered.

\begin{figure*}[t]
    \centering
    \includegraphics[width=18cm,trim=0.2cm 0.2cm 0.2cm 0.2cm,clip]{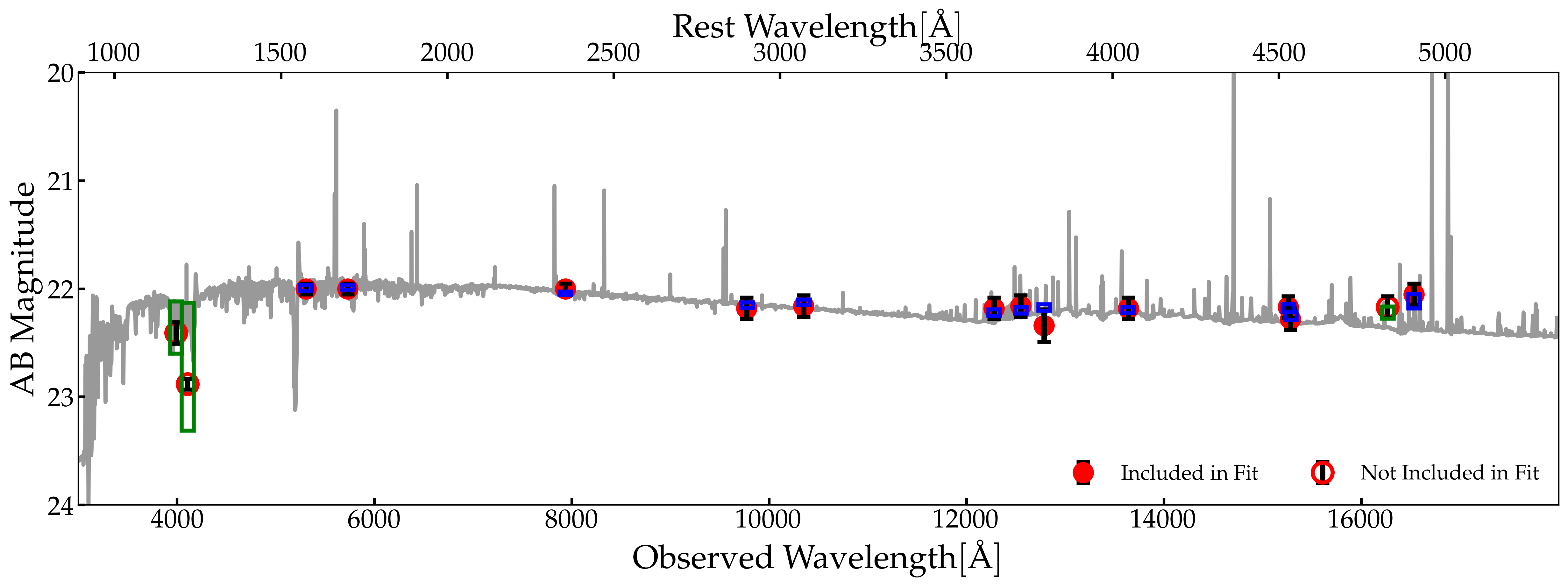}
    \caption{Best-fit SED (grey) from \texttt{Cloudy} modeling with the Chemically Anomalous Model (\refsec{chemanomaly}). Co-plotted is the 68\% C.I. range of the model predicted filter magnitude (open boxes) and the observed magnitude (circles with error bars). Two of the bluest filters F390W and F410M (open circles and green boxes) are excluded from the fit since the fluxes might have been significantly affected by Ly$\alpha$ line transfer and damping wing effects. The narrow-band F164N filter, significantly enhanced by H$\beta$, is not used due to potential flux calibration issues; however, its measured flux is consistent with model prediction.}
    \label{fig:sed}
\end{figure*}
\begin{figure*}[t]
    \centering
    \includegraphics[width=18cm]{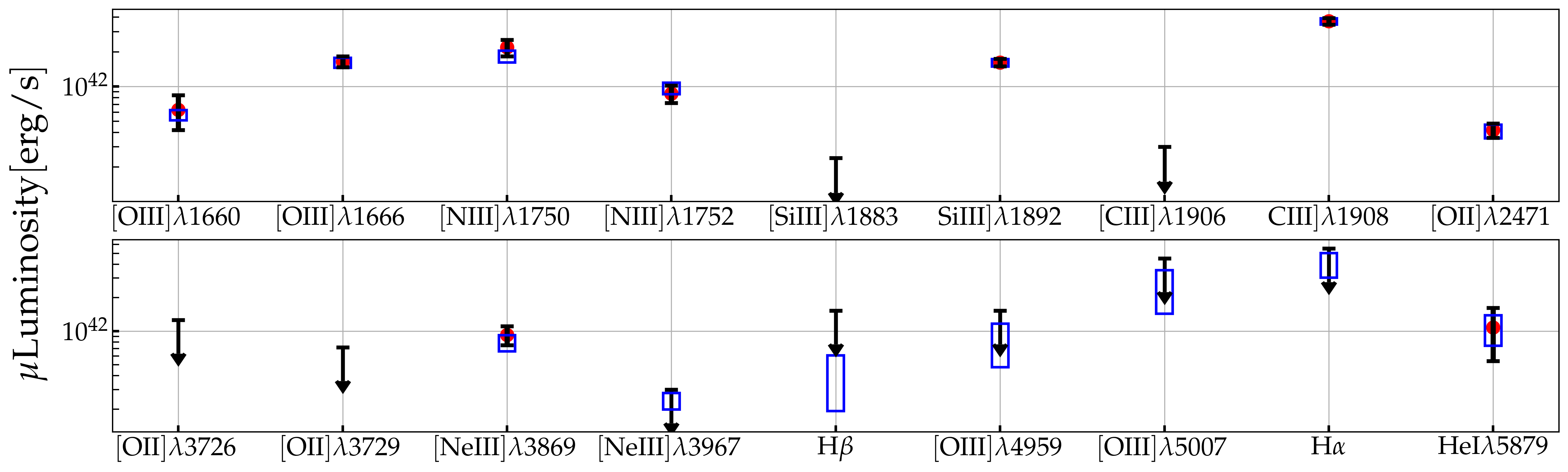}
     \caption{68\% C.I. for model predicted line fluxes (blue boxes) in the Chemically Anomalous Model (\refsec{chemanomaly}) and observed line fluxes from the MUSE and X-Shooter data (circles with errorbars). For non-detections, upper limits are plotted as downward arrows, which are implemented in fitting as strict upper limits.}
    \label{fig:lineflux}
\end{figure*}
\begin{figure}[t]
    \centering
    \includegraphics[width=\columnwidth]{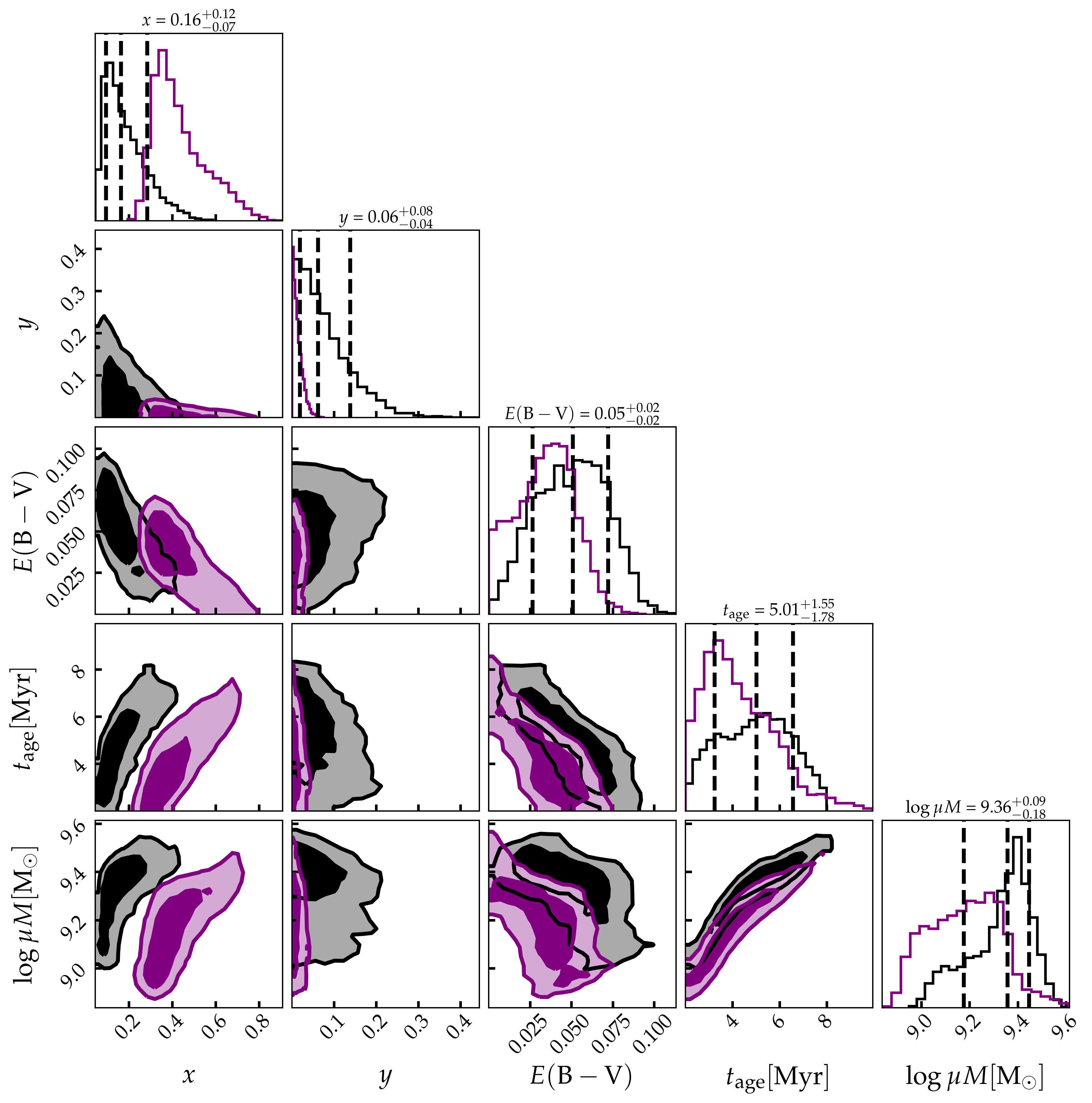}
    \caption{Posterior distributions for the star cluster parameters in the Chemically Anomalous Model (black) and the Base Model (purple). 1D histograms representing the posterior distributions are marked with the mean and 68\% confidence intervals for the Chemically Anomalous Model (C.I.'s, dashed lines). 2D contours enclose 50\% and 90\% of the posterior samples.
   Refer to Table~\ref{tab:cloudtable} for the 68\% C.I.'s of the parameters.
    }
    \label{fig:starscorner}
\end{figure}

\begin{figure}[t]
    \centering
    \includegraphics[width=\columnwidth]{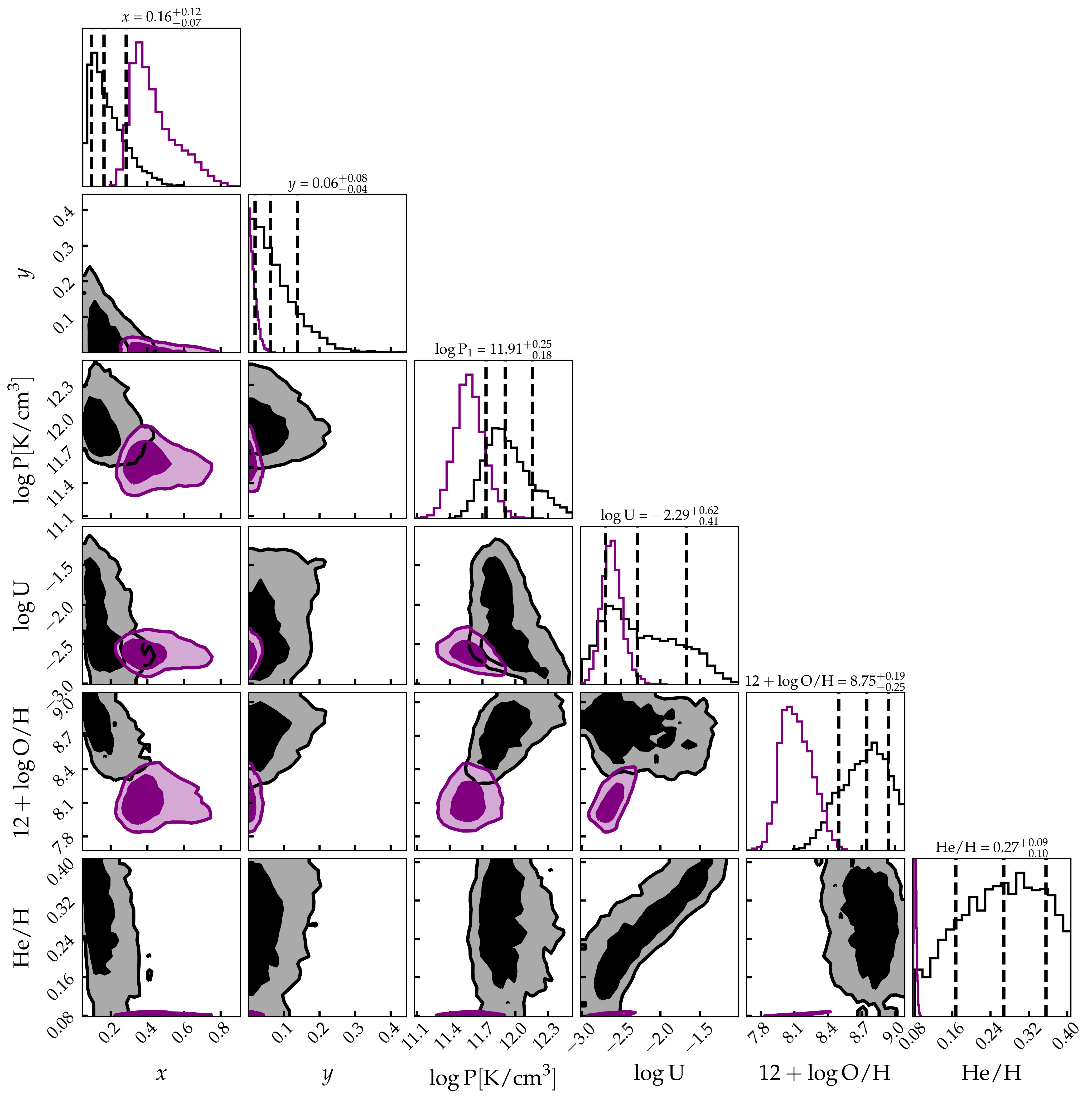}
    \caption{ Posterior distributions for the physical parameters describing the nebula for the Chemically Anomalous Model (black) and for the Base Model (purple). Contours levels and histogram confidence intervals follow Fig.~\ref{fig:starscorner}. The 68\% C.I.'s of the parameters are tabulated in  Table~\ref{tab:cloudtable}.
    }
    \label{fig:cloudcorner}
\end{figure}

\begin{figure}[t]
    \centering
    \includegraphics[width=\columnwidth]{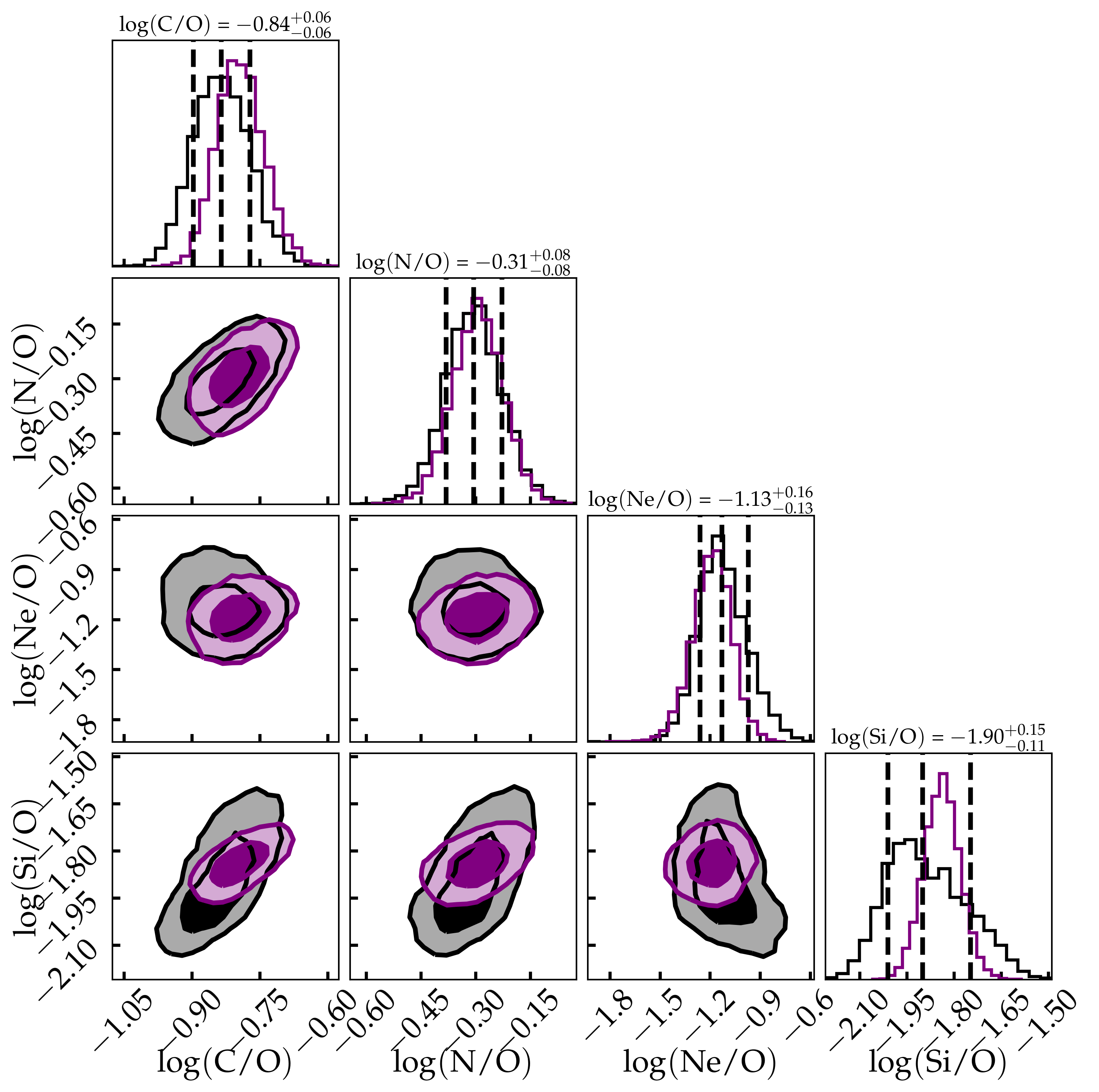}
    \caption{Posterior distributions for the gas-phase abundances of C, N, Ne and Si relative to O for the Chemically Anomalous Model (black) and the Base Model (purple). The 68\% C.I.'s of the parameters are tabulated in Table~\ref{tab:cloudtable}.
    }
    \label{fig:abuncorner}
\end{figure}

\begin{table}
	\centering
	\caption{Fitting results for the Base Model and the Chemically Anomalous Model. In the Base Model the gas-phase oxygen abundance is identified with the stellar metallicity, while in the Chemically Anomalous Model we fix the stellar metallicity to $0.25\,Z_\odot$ but allows gas-phase O and He abundances to vary in \texttt{Cloudy} calculations. Pressures are in units of ${\rm K}\,{\rm cm}^{-3}$, photon fluxes ($\Phi({\rm H}^0)$) in units of ${\rm cm}^{-2}\,{\rm s}^{-1}$, and the characteristic nebula radii ($R$) in units of ${\rm cm}$. The magnification factor of Godzilla, $\mu$, is likely large $\mu\sim 600$--$2000$ but uncertain (see \refsec{mag}).}
 	\label{tab:cloudtable}
	\begin{tabular}{c|cc}
		\hline
            \hline
        Parameter & Base Model & Chem. Model \\
        \hline
        $\chi^{2}_{\nu}$ & 0.52 & 0.39 \\
        $\chi^{2}$ & 9.36 & 6.63 \\
        \hline
        $\Ebv$ & $0.02^{+0.02}_{-0.02}$ & ${0.05}_{-0.02}^{+0.02}$ \\
        $\log Z^{a}_{\star}$ & $-2.29^{+0.16}_{-0.12}$ & -2.3 \\
        $t_{\rm age}$~[Myr] & $4.03^{+1.78}_{-1.22}$ & ${5.01}_{-1.78}^{+1.55}$\\
        $\log U$ & $-2.61^{+0.13}_{-0.011}$ & ${-2.29}_{-0.41}^{+0.62}$ \\
        $\log P$ & $11.56^{+0.14}_{-0.12}$ & ${11.91}_{-0.18}^{+0.25}$ \\
        $\log(\mu\,M_{\star})$ & $9.29^{+0.11}_{-0.16}$ & ${9.36}_{-0.18}^{+0.09}$ \\
        $\log(\mu\,Q({\rm H}_{1}^{0}))$ & ${55.54}_{-0.25}^{+0.17}$ & ${55.54}_{-0.27}^{+0.23}$ \\
        $\log \Phi({\rm H}^{0})$ & $15.22^{+0.16}_{-0.15}$ & ${15.89}_{-0.34}^{+0.38}$ \\
        $\log (R\,\mu^{1/2})$ & $19.60^{+0.12}_{-0.13}$ & ${19.25}_{-0.17}^{+0.20}$ \\
        \hline
        $x$ & $0.40^{+0.15}_{-0.09}$ & ${0.16}_{-0.07}^{+0.12}$ \\
        $y$ & $0.01^{+0.02}_{-0.01}$ & ${0.06}_{-0.04}^{+0.08}$ \\
        \hline
        $\log({\rm C/O})$ & $-0.79^{+0.05}_{-0.05}$ & ${-0.84}_{-0.06}^{+0.06}$ \\
        $\log({\rm N/O})$ & $-0.30^{+0.07}_{-0.07}$ &  ${-0.31}_{-0.08}^{+0.08}$ \\
        $\log({\rm Ne/O})$ & $-1.20^{+0.11}_{-0.12}$ & ${-1.13}_{-0.13}^{+0.16}$ \\
        $\log({\rm Si/O})$ & $-1.84^{+0.06}_{-0.05}$ & ${-1.90}_{-0.11}^{+0.15}$ \\
        $12+\log({\rm O/H})$ & $^{b}8.09^{+0.17}_{-0.11}$ & ${8.75}_{-0.25}^{+0.19}$ \\
        He/H & $^{c}{0.08}_{-0.00}^{+0.00}$ & ${0.27}_{-0.10}^{+0.09}$ \\
	\hline
        \hline
	\end{tabular}
    \tablenotetext{a}{Where $\log Z_{\odot} = -1.85$}
    \tablenotetext{b}{Derived from $Z_\star$, where $12+\log({\rm O/H})_\odot = 8.69$.}
    \tablenotetext{c}{Derived from $Z_\star$ following \cite{RussellDopita1992abunMagellanic}.}
\end{table}

\section{Discussion}
\label{sec:discuss}

In this Section, we discuss the many implications of our fitting results for the dynamics, chemical composition, and astrophysical origin of Godzilla's nebula. We will refer to our fitting results derived for the Chemically Anomalous Model as the favorable model.

\subsection{Nebular Pressure and Ionization Parameter}
\label{sec:pressure}

The ionized gas excited in Godzilla appears extremely dense $\nele \sim 10^{7-8} {\rm cm}^{-3}$ and highly pressurized $P\sim 10^{11.5-12}\,{\rm K}\,{\rm cm}^{-3}$. This result appears very robust to us. The tight lower limit on the {\rm C III]}$\lambda$1908/{\rm [C III]}$\lambda$1906 alone implies $\nele > 10^6\,{\rm cm}^{-3}$ \citep{Vanzella2020Tr}, but upper limits on {\rm [O III]}$\lambda\lambda$4959,5007 in the presence of strong {\rm [O III]}$\lambda\lambda$1650,1666 further requires $\nele > 10^7\,{\rm cm}^{-3}$ to sufficiently suppress the former by electron collision. Interestingly, the high $\nele$ reduces cooling via some collisionally excited metal lines such that dramatically increased O/H only causes a mild decrease in the electron temperature, a $\sim(10-15)\%$ drop in $T_e$ for a five-fold increase in O/H. This leads to a unique situation where, despite a solar O/H, high excitation FUV lines like {\rm [O III]}$\lambda\lambda$1660,1666 and {\rm [C III]}$\lambda\lambda$ 1906,1909 remain strong, while at typical ISM densities these lines are not observed in star-forming regions with $Z \gtrsim 0.4\,Z_\odot$~\citep{Mingozzi2024}.

\subsection{Enrichment by Massive Star Winds: N and He}

\begin{figure*}[t]
    \centering
    \includegraphics[width=18cm]{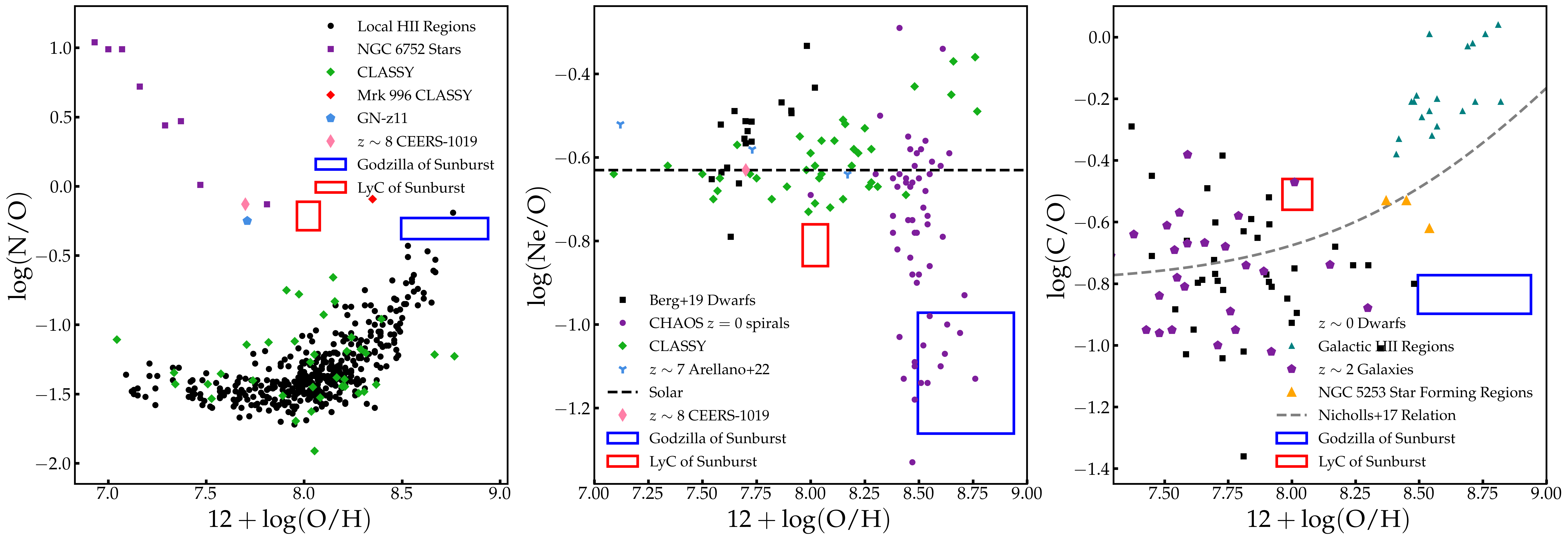}
     \caption{{\bf Left:} Nitrogen abundances for Godzilla, the LyC cluster of Sunburst~\citep{Pascale2023}, GN-z11~\citep{Senchyna2023gnz11}, the $z\sim 8$ galaxy CEERS-1019~\citep{MarquesChavez2024}, stars of local globular clusters~\cite[NGC 6752][]{Carretta2005NGC6752}, local star forming galaxies \cite[CLASSY]{Berg2022CLASSY,Stephenson2023CLASSY}, and local HII regions in SDSS galaxies (\cite{Pilyugin2012HIIRegions}). {\bf Middle:} Neon abundance of Godzilla compared to local star-forming galaxies from CLASSY and \cite{Berg2019MetalPoorCNOEvolution}, star-forming regions of M33 from \cite[CHAOS]{Rogers2022}, the $z\sim7$ galaxies of \cite{Arellano2024}, and the solar value $\log({\rm Ne/O}) = 0.63\pm 0.05$ \citep{Asplund2021SolarAbund2020}. {\bf Right:} Carbon abundance of Godzilla compared to local star-forming galaxies \citep{Berg2016,Berg2019MetalPoorCNOEvolution,Senchyna2017,LopezSanchez2007NGC5253}, galactic HII regions \citep{Arellano2020}, a sample of $z\sim2$ galaxies from the literature \citep{Pettini2000,Fosbury2003,Erb2010,Christensen2012,Bayliss2014,James2014,Stark2014,Steidel2016,Vanzella2016,Amorin2017,Berg2018,Rigby2018} and the empirical relation derived by \cite{Nicholls2017}. Figures are recreated following \cite{Senchyna2023gnz11}, \cite{Berg2019MetalPoorCNOEvolution}, and \cite{Arellano2024}, and error bars are omitted for visual clarity.}
    \label{fig:all_abuns}
\end{figure*}

The localized, elevated N/O seen in Godzilla is significantly higher than in typical ISM environments (Fig.\ref{fig:all_abuns}), including local HII regions \citep{Pilyugin2012HIIRegions} and low-$Z$ dwarf starbursts such as in the CLASSY survey \citep{Berg2022CLASSY}. However, it is consistent with the N/O found in the LyC cluster in the same galaxy~\citep{Pascale2023}, and bears similarity to a growing list of high-$z$ N emitters, such as GN-z11~\citep{Cameron2023gnz11,Senchyna2023gnz11} and CEERS-1019~\citep{MarquesChavez2024}. \citetalias{Pascale2023} propose that the N enhancement seen in the LyC cluster likely arises from CNO-processed material ejected through massive star winds. We find this to be the most plausible explanation for Godzilla as well. Apart from the winds of classical WN stars, N-rich ejecta could also be provided through slow mass loss from non-conservative mass transfer between binary massive stars~\citep{deMink2009BinaryAbundanceAnomaly}, dense equatorial winds from fast rotating massive stars~\citep{Roy2022PreSNwindsNitrogen}, continuum-driven dense winds from Very Massive Stars~\citep{Vink2023}, or even from supermassive stars that may form at the center of a dense super star cluster~\citep{Gieles2018supermassivestar, Charbonnel2023gnz11}. All these could have happened in the cluster by $4\,$Myr.

Inside a sufficiently dense cluster, even fast line-driven winds from massive stars may be retained in the cluster potential from rapid loss of kinetic energy to radiative cooling~\citep{Wunsch2011ClusterWindWithCooling, Wunsch2017RapidCoolingRHDsimulation}. Following the criteria of the semi-analytic analysis of \cite{LochhaasThompson2017Cooling}, Godzilla appears compact enough ($M_*/R > 10^5 M_\odot\,{\rm pc}^{-1}$) that winds as fast as $v_w \sim 1000\,{\rm km}\,{\rm s}^{-1}$ are expected to condense in the cluster.

Enhancement in He/H by a factor of $\sim2$--$4$ than the ISM He abundance in \cite{RussellDopita1992abunMagellanic} is another piece of evidence for retained stellar wind material. Simultaneous N and He enrichment is not surprising given the inferred cluster age $>3\,$Myr. For a smaller $t_{\rm age}$, winds from the Very Massive Stars would cause N enrichment without major He enrichment~\citep[e.g.,][]{Roy2022PreSNwindsNitrogen, Vink2023}, while moderate He enhancement is expected to accompany N enhancement in the winds of evolved He stars \citep[e.g.,][]{Meynet2006,Decressin2007}.

\subsection{Enrichment by CCSNe: C/O, Si/O, and Ne/O}
\label{sec:carbon}

Another marked event is the onset of CCSNe when the star cluster evolves beyond $3\,$Myr. 
Godzilla shows a low gas-phase $\log({\rm C/O}) = -0.8$ compared to the solar value and to HII regions at the similar O/H (approximately solar). However, this C/O value is more consistent with star-forming regions of gas-phase $Z\lesssim 0.25\,Z_\odot$ at Cosmic Noon~\citep[Fig.~\ref{fig:all_abuns};][]{Garnett1995,Berg2016,Nicholls2017, Berg2019MetalPoorCNOEvolution}. 
\cite{Berg2016} and \cite{Berg2019MetalPoorCNOEvolution} observe a trend in C/O versus O/H in local and intermediate-$z$ star-forming galaxies, which appears mostly flat at SMC-like metallicities or lower, but shows a strong upturn at higher metallicities (Figure 7 of \citealt{Berg2016} and Figure 10 of \citealt{Berg2019MetalPoorCNOEvolution}). \cite{Berg2019MetalPoorCNOEvolution} posit that this trend can be explained by a model of bursty star formation, where prompt CCSNe first enhance the gas-phase oxygen to produce a mostly flat C/O trend with metallicity.
Then, pseudo-secondary production of carbon from low-mass asymptotic giant branch (AGB) stars, together with $Z$-dependent winds of evolved stars relevant only for metallicities greater than the SMC value, is responsible for an upward trend of C/O versus O/H. This provides a plausible explanation for Godzilla, in which at an age of $\sim 4$--$6\,$Myr the first CCSNe significantly enrich the cluster gas with oxygen, leading to a low gas-phase C/O. Godzilla, however, shows enhanced O/H and hence C/H, relative to the galaxies of the \cite{Berg2019MetalPoorCNOEvolution} sample. This can be explained by the different length scales of enrichment. On the galactic scale, the CCSNe ejecta are likely diluted in the ISM, leading to sub-solar O/H values. The localized nebula gas of Godzilla may consist of, either entirely or significantly, the stellar ejecta, thus exhibiting a solar-like gas-phase O/H. Because the location of the upturn observed in the C/O versus O/H trend is set by the stellar metallicity, the gas-phase C/O of Godzilla would be consistent with the values for galaxies at sub-solar metallicities despite the inferred solar value for gas-phase O/H.

To further investigate whether the nebula of Godzilla is heavily enriched by massive star winds and CCSNe ejecta, we draw comparison to the models of \citealt{Molla2012StarClusterEjecta} (hereafter \citetalias{Molla2012StarClusterEjecta}), who calculated the element abundances for the cumulative wind and CCSN ejecta of a fiducial $M_\star=10^6\,M_\odot$ star cluster across the stage of early evolution up to an age of $20\,$Myr, assuming {\it complete} ejecta retention without mixing with the ISM.

\begin{figure}[t]
    \centering
    \includegraphics[width=8.25cm,trim=0.3cm 0.25cm 0.25cm 0.25cm,clip]{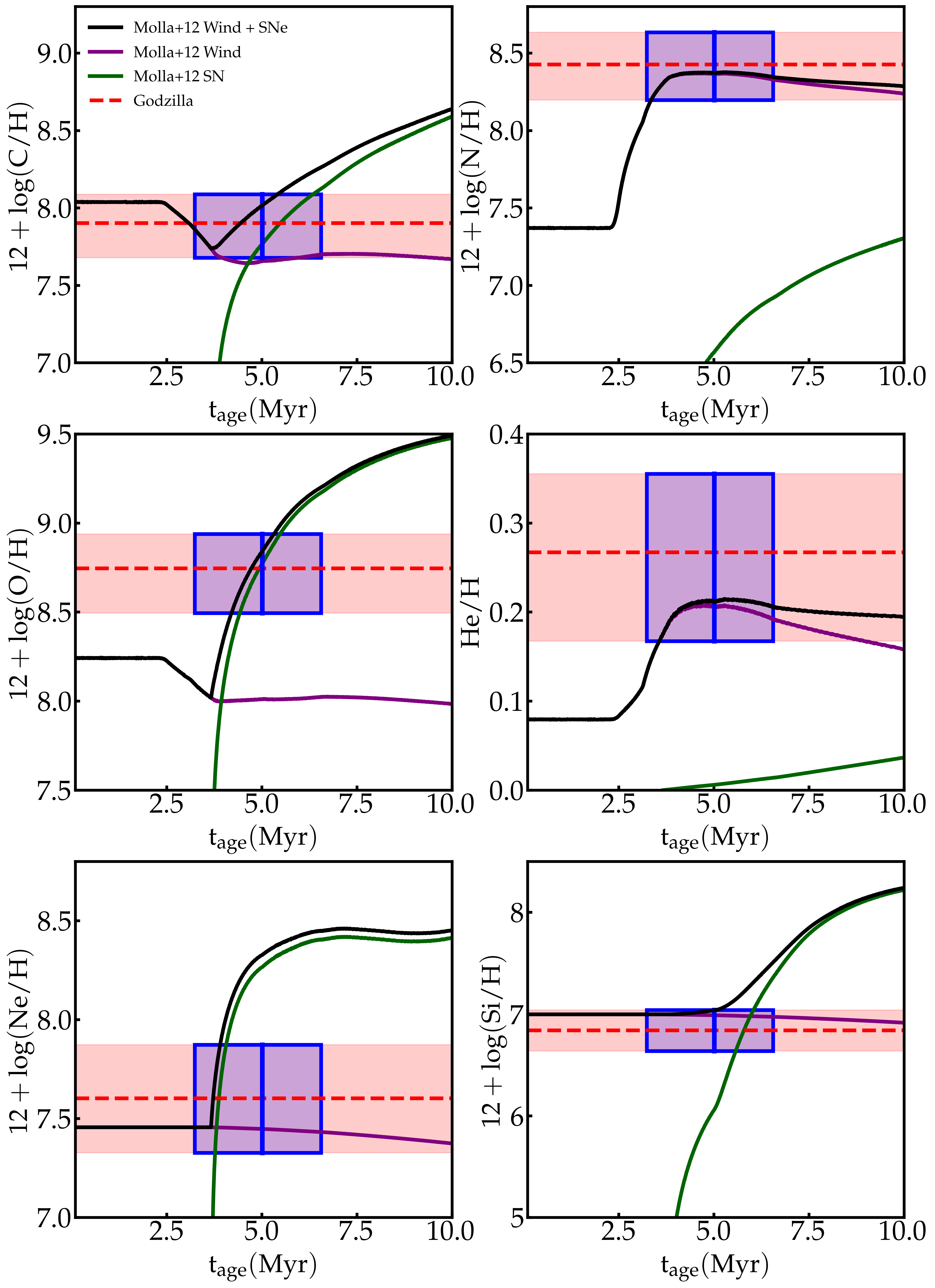}
    \caption{Evolution of the abundances of accumulated wind and CCSN ejecta. Model predictions from \citetalias{Molla2012StarClusterEjecta} (black solid lines) are compared to abundances inferred for Godzilla with the Chemically Anomalous Model (dashed red lines and red swaths represent the median and 68\% confidence interval respectively).
    The relative contributions of winds and SN to the total ejecta are plotted in purple and green respectively. We note that these represent the contribution to the abundance of the cumulative ejecta and not the abundances of the wind or SN themselves.
    For reference, we also plot for cumulative ejecta with CCSN yields of Ne and Si excluded (black dashed lines). We find the cluster age $\sim 4$--$6\,$Myr (blue box) inferred from the Chemical Anomalous Model, is in good agreement with the \citetalias{Molla2012StarClusterEjecta} predictions, such that Godzilla's nebula may be an order-unity condensation of wind and CCSN ejecta.}
    \label{fig:molla}
\end{figure}

\begin{figure}[ht]
    \centering
    \includegraphics[width=8.25cm,trim=0.3cm 0.25cm 0.25cm 0.25cm,clip]{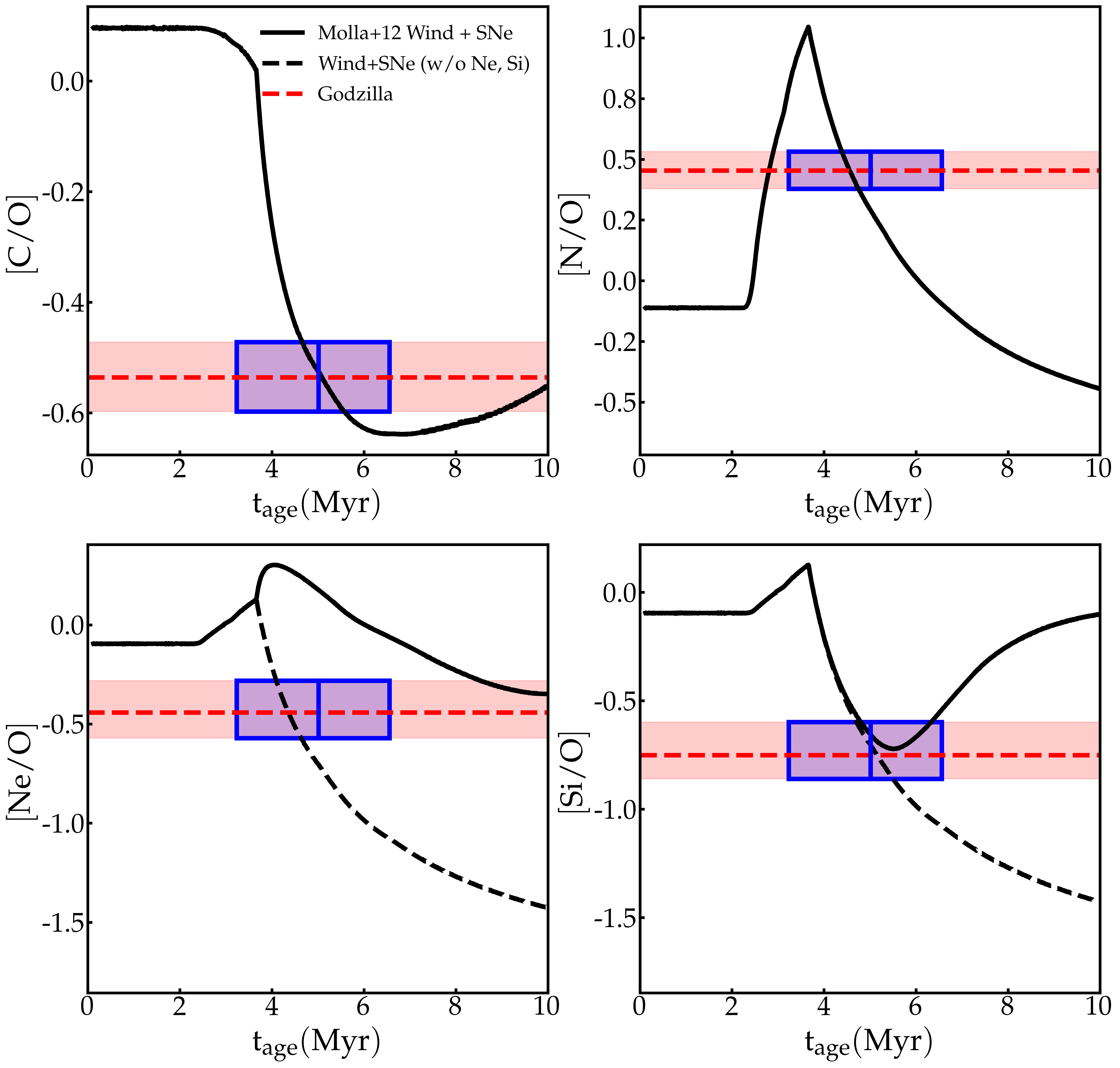}
    \caption{Same as Figure~\ref{fig:molla} but with abundances plotted relative to oxygen and normalized to the solar value. We find that the \citetalias{Molla2012StarClusterEjecta} predictions agree with the inferred abundances of Godzilla at $\sim 4$--$6\,$Myr, with the exception of Ne/O, which only shows agreement if the CCSN Ne yield is significantly reduced (dashed line).}
    \label{fig:molla_o}
\end{figure}

This is intended to be a qualitative comparison, as there are numerous differences between our cluster model for Godzilla and the models assumed in \citetalias{Molla2012StarClusterEjecta}, in addition to other underlying theoretical uncertainties. First, \citetalias{Molla2012StarClusterEjecta} used a Kroupa IMF \citep{Kroupa2001} with lower and upper mass cutoffs at $m_{\rm min} = 0.15\,M_\odot$ and $m_{\rm max} = 100\,M_\odot$ respectively, and used stellar evolution tracks that do not account for the effects of stellar rotation or binary evolution. Additionally, our Godzilla model has $Z=0.25\,Z_\odot$ for the stars, while the set provided in \citetalias{Molla2012StarClusterEjecta} with the closest stellar metallicity has $Z=0.2\,Z_\odot$. While $Z=0.2\,Z_\odot$ or $0.25\,Z_\odot$ are well within our inference uncertainties and does not significantly change the incident stellar radiation, this is in a range of transition where the abundance of classical WR stars appears to depend sensitively on $Z$. Finally, we shall assume that the Ne and Si abundances of the wind shown in Figures \ref{fig:molla} and \ref{fig:molla_o} follow from one-fifth the solar value, as \citetalias{Molla2012StarClusterEjecta} only provide Ne and Si yields from CCSNe. Hence in the wind ejecta the Ne and Si masses are set by the H mass, assuming that hydrogen in the wind material has not been significantly depleted into He through nuclear burning.

We compare our inferred abundances for Godzilla to those derived by \citetalias{Molla2012StarClusterEjecta} in Fig.~\ref{fig:molla} and Fig.~\ref{fig:molla_o}, which are shown relative to hydrogen and to oxygen, respectively. When compared to hydrogen, we find the abundances of He, C, N, O, Si and Ne are all consistent with the \citetalias{Molla2012StarClusterEjecta} predictions for complete retention of both wind and CCSN ejecta over the inferred cluster age. This supports the hypothesis that an order-unity fraction of Godzilla's nebula gas is stellar ejecta. Given the sharp upturn in oxygen abundance after the onset of CCSNe at $\sim 3.7$~Myr in the \citetalias{Molla2012StarClusterEjecta} models, many of the abundances are only consistent within a short age window of $\sim4-6$~Myr. However, this is consistent with the cluster age we infer from broad-band photometry and nebular emission lines.

Plotting the abundances relative to oxygen (Fig.~\ref{fig:molla_o}), all abundances remain in good agreement with the \citetalias{Molla2012StarClusterEjecta} predictions, with the exception of Ne/O. In a broader context, ISM Ne/O measurements in star forming regions nearly ubiquitously show the solar value or are only mildly lower \citep{Henry1999, Willner2002}. While $\log({\rm Ne/O})<-1$ was reported for a number of low ionization regions measured in the CHAOS sample of M33 \citep[see Fig.~\ref{fig:all_abuns};][]{Rogers2022}, those exhibited large measurement uncertainties and are qualitatively different environments than a young massive star cluster. 

Rather, we suggest that a sub-solar Ne/O may be typical in the early evolution of super star clusters as a consequence of CCSNe yields for progenitors of high ZAMS masses. At an inferred age $\sim 5\,$Myr, Godzilla uniquely probes CCSNe yields within only $1$--$2\,$Myr of CCSN onset when only stars $m_{\rm ZAMS} \gtrsim 30$--$40\,M_\odot$ have ended their lives~\citep{Bressan2012PARSEC}. While tables provided by \citetalias{Molla2012StarClusterEjecta} show solar or even super-solar Ne/O values at these young ages (drawing from yield models of \citealt{Woosley1993} and \citealt{Woosley1995}), Ne/O of massive stars is subject to significant theoretical uncertainty. For example, \cite{Limongi2018} predict solar and sub-solar Ne/O from massive stars at low metallicity, and the $\log({\rm Ne/O})$ values are significantly lowered below $-1$ in rotating star models, a feature not accounted for in the yield models used by \citetalias{Molla2012StarClusterEjecta}. 
More exotic sources, such as metal-free massive star hypernovae \citep{Grimmett2018Hypernovae} or high mass pair-instability SNe \citep{Heger2002PopIII} have also been predicted to produce subsolar Ne/O \citep{Ji2024MassiveStarNucleosynthesis}.


To investigate possible reasons for the reduced Ne/O, we take the \citetalias{Molla2012StarClusterEjecta} ejecta models but exclude Ne and Si yields from CCSNe. We find that this resolves the tension in Ne/O and may also improve the agreement for Si/O. This may be evidence that CCSNe of high $m_{\rm ZAMS}$ progenitors have deficient Ne/O compared to their lower $m_{\rm ZAMS}$ counterparts. While Ne and O are both abundant in the Ne-Mg-O shell of the pre-SN star, Ne may be depleted via photo-disintegration, converting to O pre-SN or during explosive nucleosynthesis~\citep{Boccioli2024}. This may also be related to a scenario in which the Ne-Mg-O and Si-S shells interior to the pre-SN star do not successfully get ejected while the more exterior C-O shell is ejected, although this may not provide enough O enhancement relative to C to match the observed sub-solar C/O value~\citep{Woosley2002}.


Overall, and despite the various measurement and theoretical uncertainties, we find that the unusual gas-phase abundances we infer for Godzilla using the Chemically Anomalous Model can be remarkably well-explained by a model in which the nebula is a full mixture of winds and CCSN ejecta from stars with $m_{\rm ZAMS} \gtrsim 30$--$40\,M_\odot$. This interpretation is consistent with the inferred cluster age $\sim 4$--$6\,$Myr.


\subsection{Nebula Geometry and Dynamics}
\label{sec:geometry}

In this Section, we discuss the nebula geometry of Godzilla and the possible underlying dynamical reasons.

\subsubsection{Spherical Shell Enclosing Stars}
\label{sec:shellmodel}

The simplest geometry would be a spherical shell of nebular clouds surrounding the entirety of the ionizing sources at some distance $R_{\rm sh}$, with an area covering factor $x \simeq 0.1$--$0.2$. In \citetalias{Pascale2023}, we found that this simple geometry can explain the spectra of the LyC cluster. For $\log P=11.9$, $\log U=-1.5$ and He/H=$0.35$ (\reffig{cloudcorner}), we require $\nele\sim 10^7\,{\rm cm}^{-3}$, and the characteristic ionizing flux is $\log\Phi({\rm H^0})=15.9$. Comparing this to the BPASS prediction of ionizing photon production rate, we obtain $\log(R\,\mu^{1/2})=19.3\,[{\rm cm}]$ or $R_{\rm sh} \simeq 0.2\,{\rm pc}\,\mu^{-1/2}_{1000}$. 
While such a compact size would be implausible for the entirety of a star cluster that is more massive than $10^6\,M_\odot$, the shell would only have to enclose the majority of the ionizing stars which may aggregate toward the cluster center~\citep{Sung2004NGC3603, Stolte2002ArchesMassFunction, Crowther2010R136VMS, Pang2013NGC3603MassSegregation}, but not necessarily all cluster stars. While it would help to place the major ionizing sources all within $\simeq 0.2\,$pc by invoking primordial mass segregation, given the large cluster mass and its age $\sim 4$--$6\,$Myr (so that the most massive stars have died), this compactness is still quite extreme.

Alternatively, the shell may have a 100\% area covering but only encloses $x=(10$--$20)\,$\% of the cluster's ionizing output. This, however, would require an even smaller shell radius $R_{\rm sh}=0.1\,{\rm pc}\,\mu^{-1/2}_{1000}\,(x/0.2)^{1/2}$. A compact gas shell enclosing a hollow cavity in the cluster center seems geometrically unnatural. We note, however, that in case the shell is thick this geometry would resemble our most favored geometry to be discussed in \refsec{core}, and the large hollow cavity could have been driven by a recent CCSN.

\subsubsection{Shattered Clouds}
\label{sec:shattered}

A second possibility is that the nebula emission comes from a population of shattered, self-shielding clouds, which are hovering within a spherical volume of radius $R_c \simeq 0.2\,{\rm pc}\,\mu^{-1/2}_{1000}$ that encloses nearly all of the cluster's ionizing output, but have an area covering factor $x\simeq 0.1$--$0.2$. The compactness of this volume may be explained again by mass segregation.

Similar to the geometry discussed in \refsec{shellmodel}, we have to explain the high pressure $\log P\simeq 11.9$ of the nebula gas. Since we inferred $\log U < -1$, direct ionizing radiation pressure is insufficient~\citep{YehMatzner2012RadiationWindFeedback},
\begin{align}
P_{\rm rad}= 6 \times 10^{10}\,{\rm K}\,{\rm cm}^{-3}\,\left(\frac{\Phi({\rm H}^0)}{10^{15.9}\,{\rm s}^{-1}\,{\rm cm}^{-2}}\right)\,\left(\frac{\langle h\nu \rangle_{\rm ion}}{20\,{\rm eV}}\right),
\end{align}
where $\langle h\nu \rangle_{\rm ion}$ is the average energy of the ionizing photons. Thus, thermal pressure from an intracloud hot gas component with $\log P_{\rm hot} \simeq 11.9$ needs to confine the clouds.

Can a global hot cluster wind provide the sufficient pressure? In a simple picture, mass and kinetic energy output from individual stellar winds and CCSNe merge to drive a hot cluster wind~\citep{Chevalier1985Nature, CantoRagaRodriguez2000ClusterWind, Wunsch2011ClusterWindWithCooling, Silich2011ClusterWindExponentialProfile}. Estimating from BPASS (v2.2), the total stellar mass loss rate is $\dot{M}_\star \approx 0.01 M_\odot\,{\rm yr}^{-1}$ per $10^6\,M_\odot$ of stars, including wind and SN mass ejection. Taking a specific CCSN rate at $t_{\rm age} \sim 5\,$Myr of $\Gamma_{\rm SN} \approx 5\times 10^{-10}\,{\rm yr}^{-1}\,M_{\odot}^{-1}$ and an explosion energy $E_{\rm SN} = 10^{51}\,{\rm erg}$ per CCSN, the mechanical luminosity due to CCSNe is $L_{\rm mech, SN} \approx 1.6 \times 10^{40}\,{\rm erg}\,{\rm s}^{-1}$ per $10^6\,M_\odot$ of stellar mass. Including wind mass loss at $v_w \approx 2000 \,{\rm km}{\rm ~s}^{-1}$ causes at most a factor of two increase in the total mechanical luminosity, reaching $L_{\rm mech, tot} \approx 3 \times 10^{40}\,{\rm erg}\,{\rm s}^{-1}$ per $10^6\,M_\odot$ of stellar mass.

Assuming no radiative loss, we find a pressure for the cluster hot gas:
\begin{align}
    & P_{\rm hot} \approx 5 \times 10^{11}\,{\rm K~cm}^{-3}\,\left( \frac{M_{\star,6}}{2} \right)\, \left(\frac{R_c}{0.2\,{\rm pc}}\right)^{-2} \nonumber\\
    & \times \left(\frac{L_{\rm mech, tot}}{3\times 10^{40}\,{\rm erg}\,{\rm ~s}^{-1}}\right)^{\frac{1}{2}} \, \left(\frac{\dot{M}_\star}{0.01\,M_\odot\,{\rm yr}^{-1}}\right),
\end{align}
where $M_{\star,6}=M_\star/(10^6\,M_\odot)$ and $R_c$ is the radius of the cluster core region where the hot gas is fully thermalized. We obtain a sufficiently high value because we have used a compact radius $R_c=0.2\,$pc for thermalization, which we deem quite extreme. 

However, we emphasize that the above analysis is very optimistic because it assumes maximal thermalization without radiative loss. Any radiative loss lowers the hot gas temperature and the pressure. Taking $\log P_{\rm hot}=11.9$, the hot gas needs to have a high density $n_{\rm hot}=8 \times 10^4\,{\rm cm}^{-3}\,(10^7\,{\rm K}/T_{\rm hot})$ and hence a short cooling time $t_{\rm cool}\simeq 60\,{\rm yr}\,(T_{\rm hot}/10^7\,{\rm K})^{1.7}\,(n_{\rm hot}/10^5\,{\rm cm}^{-3})$~\citep{Draine2011ISMtext}. This is not longer than the timescale of streaming across the core region $t_{\rm cr}=R_c/v_{\rm hot}=200\,{\rm yr}\,(R_c/0.2\,{\rm pc})/(v_{\rm hot}/1000\,{\rm km}\,{\rm s}^{-1})$, and is shorter than the typical time separation between successive CCSNe $\sim 1/(\Gamma_{\rm SN}\,M) \simeq 900\,{\rm yr}\,(\Gamma_{\rm SN} / 5\times 10^{-10}\,{\rm yr}^{-1}\,{\rm M_\odot}^{-1})^{-1}\,\left(M/2\times 10^6\,M_\odot\right)^{-1}$, even if all CCSNe occur at the cluster center. Therefore, such a compact hot cluster gas component seems unstable to radiative cooling.

Importantly, we expect $y \simeq 1/2$ for the shattered-cloud geometry, since the dense clouds would naturally have a more or less isotropic distribution about the cluster center. A $y$ value that strongly hints at the HI-obscured geometry, $y<0.2$, does not favor this geometry.


\subsubsection{HII Bubbles within Condensed Gas}
\label{sec:core}

Following previous discussions, dense cluster hot gas inevitably enters the regime of catastrophic radiative cooling~\citep{Silich2004RadiativeClusterWindS, Wunsch2011ClusterWindWithCooling, Gray2019SuperwindCoolingI, Danehkar2021SuperwindCoolingII, LochhaasThompson2017Cooling, Wunsch2017RapidCoolingRHDsimulation}, so that it would fail to blow out at all. It is found in hydrodynamic simulations that in this regime, gas undergoes rapid radiative cooling and accumulates at the cluster center~\citep{Wunsch2017RapidCoolingRHDsimulation}.

Motivated by this insight, we propose a different nebula geometry (\reffig{cartoon}): catastrophically cooled cluster gas expelled from the massive stars has accumulated in the cluster core since the beginning of the cluster evolution. This gas is dense enough that it self-shields against ionizing photons even though it is interspersed between the O stars. If the gas is not dusty (no significant dust reddening is seen toward Godzilla), its neutral interior can still be heated due to photoionization of metals by the intense UV light and hence stay warm at a few thousand Kelvin. The nebular emission may then come from ionization-bounded HII bubbles surrounding individual O stars which are embedded in self-shielding cluster gas that fills the cluster core. 

This scenario is motivated by several hints. First, special chemical abundances are indicative of a nebula gas origin from retained wind and CCSN ejecta; catastrophic cooling may first cause the wind material to accumulate in the cluster core, while the later CCSN ejecta may be trapped within.
Second, individual ionization-bounded HII bubbles can be very compact, which allows for high $\Phi({\rm H}^0)$ values, as we will show. Finally, fitting preference for a small $y$ value helps explain the weakness of H Balmer lines and indicates a high fraction of nebula emission that is affected by significant HI columns along the line of sight.

Each individual O star with wind mass loss rate $\dot M_w$, wind velocity $v_w$ and ionizing photon rate $S_0$ blows a bubble around it within the dense cluster gas. The bubble feels a ``headwind'' of the ambient cluster gas due to the relative velocity between the star and the ambient gas flow. With efficient radiative cooling, shell gas can form from this interaction~\citep{Scherer2016AstrosphereShockStructure, Titova_2021}.
Accounting for wind and radiation pressures, the bubble radius $R_b$ along the stagnation line can be estimated from $P=(4\,\dot M_w\,v_w + \Phi_0\,\langle h\nu \rangle_{\rm ion})/(4\pi\,R_b^2)$, where $\langle h\nu \rangle_{\rm ion} \approx 20\,$eV is the typical energy of the ionizing photons and a factor of 4 in front of $\dot M_w\,v_w$ accounts for the thickness of the shocked wind. The bubble has a compact size:
\begin{align}
\label{eq:Rb}
    & R_b = 170\,{\rm AU}\,\left(\frac{1 + \zeta}{2.5} \right)^{\frac12}\,\left(\frac{S_0}{10^{48.5}\,{\rm s}^{-1}}\right)^{\frac12}\,\left(\frac{\langle h\nu \rangle_{\rm ion}}{20\,{\rm eV}}\right)^{\frac12}\nonumber\\
    & \times \left(\frac{P}{10^{11.9}\,{\rm K}\,{\rm cm}^{-3}}\right)^{-\frac12},
\end{align}
where we define $\zeta = (4 \dot M_w\,v_w)/(S_0\,\langle h\nu \rangle_{\rm ion})$. The fiducial values, $S_0=10^{48.5}\,{\rm s}^{-1}$, $\dot M_w=10^{-7}\,M_\odot\,{\rm yr}^{-1}$, $v_w=2000\,{\rm km}\,{\rm s}^{-1}$ and hence $\zeta=1.5$, are chosen for median O stars that contribute to the ionizing output at $t_{\rm age}=5\,$Myr ($m_{\rm ZAMS}\simeq 25\,M_\odot$)~\citep{Bressan2012PARSEC}. If instead $t_{\rm age}=4\,$Myr, we have $S_0=10^{49}\,{\rm s}^{-1}$, $\dot M_w=10^{-6.5}\,M_\odot\,{\rm yr}^{-1}$ but again with $\zeta=1.5$, for median O stars with $m_{\rm ZAMS}\simeq 40\,M_\odot$, which corresponds to $R_b=300\,$AU. Since the HII bubbles are much smaller than the typical separation between O stars within the cluster, their total volume filling fraction is tiny.

In the ionized gas where nebular emission lines form, we have $P\approx 2\,\nele\,\kB\,\Te$. For the ionizing photon rate $S_0$ from one O star, the ionization parameter is $U = S_0/(4\pi\,R_b^2\,\nele\,c)$. This gives $U \approx (2\,\kB\,\Te)/(\langle h\nu \rangle_{\rm ion}\,(1+\zeta))$, or numerically
\begin{align}
    U = & 10^{-1.5}\,\left(\frac{1+\zeta}{2.5}\right)^{-1}\,\left(\frac{\Te}{10^4\,{\rm K}}\right)\,\left(\frac{\langle h\nu \rangle_{\rm ion}}{20\,{\rm eV}}\right)^{-1},
\end{align}
which is insensitive to the exact value of the ambient pressure. This $\log U$ value is higher than what is inferred from the Base Model, but aligns more with solutions of elevated He/H found with the Chemical Anomalous Model (\reffig{cloudcorner}). There is uncertainty in this estimation due to the anisotropic structure of the HII bubble given the motion of the O star relative to the cluster gas (e.g., \citealt{Mackey2023IAUWindInteractionISM}). We also caution about the uncertainty that the hot ionized bubbles could be larger with a lower $\log U$ value. It may be that some of these large bubbles are driven by ongoing SNRs from recent CCSNe (invoked to drive supersonic turbulence), a high-density, scaled-down analog of what happens to the multi-phase ISM in the Galaxy~\citep{McKeeOstriker1977ISMSNfeedback}.

We suggest that the stellar ejecta accumulated in the cluster core is in a state of supersonic turbulence, analogous to giant molecular clouds. Maintained by UV heating, the gas is only at a few thousand K, and its sound speed $c_s$, at several ${\rm km}\,{\rm s}^{-1}$, is smaller than the virial velocity in the cluster potential, which easily reaches tens of ${\rm km}\,{\rm s}^{-1}$ for $M_\star\sim 10^6\,M_\odot$ and $R\sim 1\,$pc. Supersonic turbulence is therefore required to support this gas, at least for one or several Myr, against gravitational collapse which would quickly turn it into stars. The turbulence velocity should be on the order of the virial velocity in the cluster core. 

Most of the volume is filled with the turbulent HI gas. For a large turbulence Mach number $\mathcal{M}=u/c_s$ (where $u$ is the turbulence velocity on the largest scale), the gas is highly inhomogeneous~\citep{Hennebelle2008AnalyticTheoryIMF, Hopkins2012ExcursionSets}. The individual HII bubbles around O stars mostly probe the volume-weighted median density $n_V$, which should be high enough to provide the ram pressure in the ``headwind'':
\begin{align}
\label{eq:nV}
    n_V \simeq & \frac{P}{m_{\rm p}\,u^2} = 3 \times 10^6\,{\rm cm}^{-3} \nonumber \\
    & \times \left(\frac{P}{10^{11.9}\,{\rm K}\,{\rm cm}^{-3}}\right)\,\left(\frac{u}{50\,{\rm km}\,{\rm s}^{-2}}\right)^{-2}.
\end{align}

The typical HI column density in front of the nebular source is $N_{\rm HI} \simeq n_V\,R_g =2\times 10^{24}\,{\rm cm}^{-2}\,(n_V / 10^{6.5}\,{\rm cm}^{-3})\,(R_g / 0.2\,{\rm pc})$. Viewing from the outside, the FUV sources inside the HI gas is analogous of an extremely damped Ly$\alpha$ system. In fact, for $N_{\rm HI}=2\times 10^{24}\,{\rm cm}^{-2}$, the Ly$\alpha$ damping wing absorption equivalent width is huge $\sim 1000\,\AA$, which means FUV flux in the wavelength range of H Lyman series should be completely attenuated. However, this does not necessarily contradict data, as in our model FUV sources embedded within the retained HI gas make up a small fraction (e.g. $\lesssim 10\%$) of the cluster's total FUV flux. A lot more FUV sources are expected to be exterior to the HI gas; ionizing radiation are expected to be more concentrated than FUV radiation in dense young star clusters~\citep{Kim2023}. 

A more important, fundamental difference between damped Ly$\alpha$ systems and the geometry we consider is that for the former photons are scattered out of line of sight, while for Godzilla photons do have to eventually emerge from the HI gas cloud somewhere. Accounting for photons scattered back into the line of sight therefore reduces the trough width. Indeed, F390W and F410M fluxes appear to be moderately deficient but far from being completely attenuated (\reffig{sed}).

According to the theory of supersonic turbulence, the mean density $\overline{n}$ should be larger than $n_V$ by a factor $e^{\sigma^2_\delta/2}$~\citep{Hennebelle2008AnalyticTheoryIMF, Padoan2014StarFormationTheoryReview}, where $\sigma^2_\delta\simeq 4$--$5$ for $\mathcal{M} \simeq 10$~\citep{Federrath2021NatSonicScaleTurbulence}. Thus, $\overline{n}$ should be a factor of $10$ larger than \refeq{nV}.

The retained gas core radius $R_g$ may be a fraction of the cluster size since we infer a small covering factor $x \simeq 0.2$. The gas mass required is
\begin{align}
    M_g \approx 3\times 10^4\,M_\odot\,\left(1 + \frac{4\,{\rm He}}{{\rm H}}\right)\,\left(\frac{\overline{n}}{10^{7.5}\,{\rm cm}^{-3}}\right)\,\left(\frac{R_g}{0.2\,{\rm pc}}\right)^3.
\end{align}
For a cluster mass $M_\star = 2\times 10^6\,M_\odot/\mu_{1000}$ and if $R_g$ is a small fraction of a parsec, there is sufficient wind mass loss and CCSN ejecta to account for $M_g \lesssim 3 \times 10^4\,M_\odot$.

If we identify the turbulence outer scale with the radial extent of the accumulated gas cloud $R_g$, the sonic scale~\citep{Padoan1995SupersonicTurbulenceFragmentation},
\begin{align}
    l_s = \left( \frac{c_s}{u} \right)^2\,R_c = 270\,{\rm AU}\,\left(\frac{R_g}{0.2\,{\rm pc}}\right)^2\,\left(\frac{\mathcal{M}}{10}\right)^{-2},
\end{align}
is larger or comparable to the typical HII bubble size $R_b$ (see \refeq{Rb}). Hence the ``headwind'' does not behave clumpy before colliding with the O star wind.

The SNR expands into the cluster gas at the volume-weighted median density $n_V$. For $n_V \gtrsim 10^6\,{\rm cm}^{-3}$, the SNR can fail to break out of the condensed cluster gas before it dissipates, even for $R_g$ as small as a fraction of a parsec. This allows the condensed gas to stay in the cluster potential, and injects a small fraction of the explosion energy into the gas to maintain turbulence. We will discuss this in \refsec{driving}.

We note that the geometry of individual HII bubbles embedded within turbulent HI gas can explain the preference for a small $y$ value. There should still be ionizing sources beyond the radius $R_g$, which illuminate the cluster gas cloud from the outside. A $y$ value close but not equal to zero may account for such non-HI-obscured contributions. This is an advantage of this model compared to the scenario of shattered dense clouds with a low area covering factor. We believe this unusual geometry best describes Godzilla in reality.

\subsection{SN Driving of Turbulence}
\label{sec:driving}

The state of supersonic turbulence is highly dissipative, radiating away the bulk of the flow energy on the order of the turnover time of the largest eddy, which should be on the order of the dynamic time (which is much shorter than $\sim$Myr). The energy dissipation rate is estimated to be
\begin{align}
\label{eq:dEdt_tb}
    & \left( \frac{\rmd E}{\rmd t} \right)_{\rm turb} \sim \frac{u^2}{2}\,\left(1 + \frac{4\,{\rm He}}{{\rm H}}\right)\,m_p\,\overline{n}\,\frac{4\pi}{3}\,R^3_g\,\left( \frac{R_g}{u} \right)^{-1} \nonumber\\
    & \simeq 4\times 10^{39}\,{\rm erg}\,{\rm s}^{-1}\,\left(1 + \frac{4\,{\rm He}}{{\rm H}}\right)\,\left(\frac{\overline{n}}{10^{7.5}\,{\rm cm}^{-3}}\right) \nonumber\\
    & \times \left(\frac{R_g}{0.2\,{\rm pc}}\right)^2\,\left(\frac{u}{50\,{\rm km}\,{\rm s}^{-1}}\right)^3.
\end{align}

We surmise that supersonic turbulence can be sustained by CCSNe after $t_{\rm age}\simeq 3\,$Myr, at least when all CCSNe occur well within the retained gas. The bulk of the CCSN explosion energy will be radiated away, but a subdominant fraction of it can maintain gas turbulence.

Let us crudely estimate CCSN driving using the semi-analytic theory of a spherical symmetric SN remnant expanding into an ambient gas of uniform density $n_V$. The initial adiabatic expansion~\citep{Sedov1959book, Taylor1950blastwave} creates a growing hot bubble of size as a function of time $t$
\begin{align}
R_{\rm SNR} \simeq 0.04\,{\rm pc}\,\left(\frac{E_{\rm SN}}{10^{51}\,{\rm erg}}\right)^{\frac15}\,\left(\frac{n_V}{10^{6.5}\,{\rm cm}^{-3}}\right)^{-\frac15}\,\left(\frac{t}{10\,{\rm yr}}\right)^{\frac25}.
\end{align}
A homologous Sedov-Taylor bubble has a temperature
\begin{align}
    T \simeq 6\times 10^6\,{\rm K}\,\left(\frac{E_{\rm SN}}{10^{51}\,{\rm erg}}\right)^{\frac25}\,\left(\frac{n_V}{10^{6.5}\,{\rm cm}^{-3}}\right)^{-\frac25}\,\left(\frac{t}{10^2\,{\rm yr}}\right)^{-\frac65}.
\end{align}

Shortly after the hot bubble cools radiatively, a mass shell forms at the edge of the bubble. The evolution transitions to a pressure driven phase~\citep{Cox1972SNR, Chevalier1974SNRevolution}. To analytically estimate this transition point, we use a simple cooling function for $\Lambda(T)=1.6\times 10^{-19}\,\xi_m\,(T/{\rm K})^{-1/2}\,{\rm erg}\,{\rm cm}^3\,{\rm s}^{-1}$, valid for the hot gas in the temperature range $10^5 < T/{\rm K} < 10^{7.5}$. Here $\xi_m$ is an order-unity parameter depending on the gas metallicity and $\xi_m \simeq 1$ at solar metallicity. For high $n_V$, the pressure driven phase is quickly reached at a time
\begin{align}
    t_{\rm PDS} \simeq 3\,{\rm yr}\,\left(\frac{E_{\rm SN}}{10^{51}\,{\rm erg}}\right)^{\frac{3}{14}}\,\left(\frac{n_V}{10^{6.5}\,{\rm cm}^{-3}}\right)^{-\frac{4}{7}}\,\xi^{-\frac{5}{14}}_m,
\end{align}
with a shell radius
\begin{align}
    R_{\rm PDS} \simeq 0.02\,{\rm pc}\,\left(\frac{E_{\rm SN}}{10^{51}\,{\rm erg}}\right)^{\frac{2}{7}}\,\left(\frac{n_V}{10^{6.5}\,{\rm cm}^{-3}}\right)^{-\frac{3}{7}}\,\xi^{-\frac{1}{7}}_m.
\end{align}
The shell has swept up ambient gas of mass $M_{\rm PDS}\simeq (4\pi/3)\,m_p\,n_V\,R^3_{\rm PDS}$, which is
\begin{align}
    M_{\rm PDS} \simeq 6\,M_\odot\,\left(\frac{E_{\rm SN}}{10^{51}\,{\rm erg}}\right)^{\frac{6}{7}}\,\left(\frac{n_V}{10^{6.5}\,{\rm cm}^{-3}}\right)^{-\frac{2}{7}}\,\xi^{-\frac{3}{7}}_m.
\end{align}
The shell expands at a velocity
\begin{align}
    & v_{\rm PDS} = \frac{2\,R_{\rm PDS}}{5\,t_{\rm PDS}} \\
    & \simeq 3500\,{\rm km}\,{\rm s}^{-1}\,\left(\frac{E_{\rm SN}}{10^{51}\,{\rm erg}}\right)^{\frac{1}{14}}\,\left(\frac{n_V}{10^{6.5}\,{\rm cm}^{-3}}\right)^{\frac{1}{7}}\,\xi^{\frac{3}{14}}_m. \nonumber
\end{align}
At this point, the shell carries radial momentum~\citep{OstrikerMcKee1988BlastwavesReview}
\begin{align}
    & p_{\rm PDS} = M_{\rm PDS}\,\frac{9}{14}\,v_{\rm PDS} \simeq 1.4\times 10^4\,{\rm M_\odot}\,{\rm km}\,{\rm s}^{-1}\\
    & \times \left(\frac{E_{\rm SN}}{10^{51}\,{\rm erg}}\right)^{\frac{13}{14}}\,\left(\frac{n_V}{10^{6.5}\,{\rm cm}^{-3}}\right)^{-\frac{1}{7}}\,\xi^{-\frac{3}{14}}_m. \nonumber
\end{align}
From this point onward, the SN remnant will expand substantially until the velocity of the wrinkled gas shell decreases to be comparable to the ambient turbulence velocity $u$. This process is close to but not perfectly momentum conserving. The final momentum injected into the ambient gas, $p_{\rm SN}$, is estimated to be only a few times larger than $p_{\rm PDS}$, $\epsilon_p=p_{\rm SN}/p_{\rm PDS}=2$--$3$, the remnant radius will have expanded by several fold $\epsilon_R=R_{\rm SN}/R_{\rm PDS}=2$--$5$, and will have swept up a total mass $M_{\rm SN}=\epsilon^3_R\,M_{\rm PDS}$~\citep{Cioffi1988RadiativeSNR}. The ratios $\epsilon_p$ and $\epsilon_R$ are expected to depend rather weakly on $n_V$~\citep{Cioffi1988RadiativeSNR}. 

The final kinetic energy injected into the ambient gas can be estimated as 
\begin{align}
    & K_{\rm SN} = \frac{p^2_{\rm SN}}{2\,M_{\rm SN}} = \frac{p^2_{\rm PDS}}{2\,M_{\rm PDS}}\,\frac{\epsilon^2_p}{\epsilon^3_R} \\
    & \simeq 1 \times 10^{50}\,{\rm erg}\,\left(\frac{E_{\rm SN}}{10^{51}\,{\rm erg}}\right)\,\left(\frac{\epsilon_p}{3}\right)^2\,\left(\frac{\epsilon_R}{3}\right)^{-3}. \nonumber
\end{align}
Assuming all CCSNe occur near the cluster center within the retained cluster gas around $t_{\rm age}=4$--$6\,$Myr, the total CCSN driving energy rate is
\begin{align}
\label{eq:dEdt_sn}
    & \left( \frac{\rmd E}{\rmd t} \right)_{\rm SN} = \Gamma_{\rm SN}\,K_{\rm SN}\,M_\star \nonumber\\
    & \simeq 3 \times 10^{39}\,{\rm erg}\,{\rm s}^{-1}\,\left(\frac{\Gamma_{\rm SN}}{5 \times 10^{-10}\,{\rm yr}^{-1}\,M^{-1}_\odot}\right)\,\left(\frac{K_{\rm SN}}{10^{50}\,{\rm erg}}\right)\nonumber\\
    &\times \left(\frac{M_\star}{2 \times 10^6\,M_\odot}\right).
\end{align}
Despite much complexity in estimating CCSNe kinetic feedback in the radiative regime and with possibly inhomogeneous surrounding medium~\citep{Cowie1981SNRinhomogeneous}, the order-of-magnitude agreement between \refeq{dEdt_sn} and \refeq{dEdt_tb} following simple estimations suggests that turbulence driving by CCSNe is at least plausible, provided that the cloud size $R_g$ is not much larger than a fraction of pc and the turbulence velocity does not exceed $u = 100\,{\rm km}{\rm s}^{-1}$. 

We note that having $\epsilon_R=2$--$5$ puts $R_{\rm SN}=0.04$--$0.1\,$pc for the fiducial values of $E_{\rm SN}$ and $n_V$, well within the size of the retained cluster gas $R_{\rm SN} < R_g$. This is compatible with our physical picture here that CCSNe fail to remove the retained cluster gas, hence the efficient retention of their metal yield. SN remnant completely dissipates in the surrounding turbulent gas at the time $t_{\rm SN}=\epsilon_t\,t_{\rm PDS}$, where $\epsilon_t$ can be estimated
\begin{align}
    & \epsilon_t \simeq 300\,\left(\frac{u}{50\,{\rm km}\,{\rm s}^{-1}}\right)^{-\frac{10}{7}}\,\left(\frac{E_{\rm SN}}{10^{51}\,{\rm erg}}\right)^{\frac{5}{49}}\nonumber\\
    & \times \left(\frac{n_V}{10^{6.5}\,{\rm cm}^{-3}}\right)^{\frac{10}{49}}\,\xi^{-\frac{15}{49}}_m.
\end{align}
For our fiducial model parameters, the lifetime of each SN remnant is on the order of $t_{\rm SN} \sim 10^3\,$yr and is comparable to the inverse frequency of CCSN occurrence $1/(\Gamma_{\rm SN}\,M_\star)$. If CCSNe occur too frequently compared to the remnant lifetime, complication can arise due to clustered feedback from multiple SNe, which should enhance SN momentum feedback~\citep{Gentry2019ClusteredSNfeedback3D}.

\subsection{Effect of HI Column}
\label{sec:yeffect}

Our picture of HII bubbles embedded in condensed, self-shielding cluster gas requires that emission lines traverse a large HI column before they leave the cluster gas. Intriguing, our emission line fitting favors a $y$ value close to zero, which means we are mostly seeing emission line photons propagating through the self-shielded HI column rather than directly from the irradiated faces. According to \texttt{Cloudy}, we find this optical depth effect significantly decrease H Balmer lines for $y=0$ but does not impact the metal (semi-)forbidden lines. 

\cite{FerlandNetzer1979LineTransferAGNnovae} studied line trapping of H Lyman and Balmer photons in an ionization-bounded HII region. In particular, trapped Ly$\alpha$ photons excite an $n=2$ population which can provide a substantial resonant scattering optical depth to H Balmer photons, particularly in the absence of internal dust. As \cite{Grandi1980OI8446SeyfertI} explains (also see \cite{Hamann2012EtaCarInnerEjecta}), the trapped H$\alpha$ photons can convert to Ly$\beta$ photons, which can pump the abundant O I ions in the neutral zone. H$\alpha$ can thus be lost when there is a substantial line optical depth to H Balmer photons.

Interestingly, we see a significant detection of {\rm O I}$\lambda$1641 (the line center is inconsistent with {\rm He II}$\lambda$1640), which is a fluorescent signature of Ly$\beta$ pumping of O I. We refrain from quantitatively comparing the {\rm O I}$\lambda$1641 detection to \texttt{Cloudy} predictions, as the line trapping effect is sensitive to the column density, kinematics of the neutral gas behind the ionization front, and internal grain abundance. 

This provides a promising explanation, which is perhaps more important than merely the elevated O and He abundances, for the surprising weakness of H Balmer lines as first pointed out by \cite{Vanzella2020Tr}. We take this strong preference in data for an optical depth effect from a H I column as one more supporting evidence for our physical model for Godzilla's nebular emission.

We also note that this optical depth effect results in a model-predicted unreddened H$\alpha$/H$\beta$ ratio much larger ($\sim 10$) than the standard assumption for Case B recombination \citep[$\sim 2.81$ for $n_e\sim10^6\,{\rm cm}^{-3}$;][]{OsterbrockFerland1989text}. \cite{Scarlata2024} argue that in this regime of a large excited an $n=2$ population, a spherically symmetric cloud geometry may result in high intrinsic H$\alpha$/H$\beta$ ratio due to Balmer self-absorption, while highly non-symmetric geometries could preferentially scatter $H\alpha$ photons out of the line of sight, leading to lower intrinsic Balmer decrements. This implies that using Case B assumptions for the measurement of external dust reddening may be inappropriate in this regime. In the case of our model-predictions, it would mimic a substantial extinction ($\Ebv \sim 1$) which would be inconsistent with the detection of a bright FUV stellar continuum with a blue SED slope and strong UV nebular emission lines.

\subsection{Expectations for Emission Line Profile}
\label{sec:kine}

The physical picture presented here may give some insight into understanding the emission line profile. \cite{Choe2024} show from high resolution JWST NIRSpec/IFU spectroscopy the superposition of narrow (FWHM$\lesssim 100$~km/s), broad (FWHM$\sim200$-$300$~km/s), and very broad (in the case of H$\alpha$; FWHM$\sim 500$~km/s) components to optical emission lines. While a quantitative prediction of the emission line profile and relative component strengths from our model requires extensive theoretical exploration that is beyond the scope of this work, the existence of a broad component can be qualitatively compatible with our physical picture.

In the picture of catastrophic cooling of a global cluster wind, the entire volume of the cluster is not expected to cool, except in the regime of high mass loading of the wind material with ambient cluster gas \citep{LochhaasThompson2017Cooling}. Rather, there exists a cooling radius, outside of which massive stars may still inject mass and kinetic energy to drive a low-density hot cluster wind. At intermediate radii, there is likely a multi-phased transition regime between the fully condensed, self-shielding phase at the cluster center and the fast, radiatively inefficient hot wind outside \citep{Wunsch2017RapidCoolingRHDsimulation}. There, warm gas cloudlets, optically thick to ionizing photons but with a low total area covering fraction, may be accelerated to hundreds of km/s by gas pressure or radiation, outflowing out of the cluster potential. Nebular emissions from these outflowing cloudlets, powered by stars outside the central condensed gas component, would naturally result in broader nebular emission lines compared to emission from the central region.

In principle, accurate modeling would require including nebular emission from outflowing gas cloudlets. \cite{Choe2024} found that the broad component dominates the line flux in H$\alpha$, H$\beta$ and {\rm [O III]}$\lambda\lambda$4959,5007, but in our model these are also the lines that are predicted to be unusually weak from dense gas settled at the cluster center (radiation transfer suppression for H Balmer lines and collisional suppression for {\rm [O III]}$\lambda\lambda$4959,5007). In other major optical lines, the broad component appears subdominant compared to the narrow component. The broad component appears weaker than the narrow component in aurora lines {\rm [O III]}$\lambda$4363 and {\rm [N II]}$\lambda$5755, which have critical densities around $3\times 10^7\,{\rm cm}^{-3}$. This supports the interpretation that the broad component appears dominant in {\rm [O III]}$\lambda\lambda$4959,5007 and H$\alpha$, H$\beta$ because the narrow component, which as we hypothesize comes primarily from the condensed gas settled down at the cluster center, is weak in these lines.

This is compatible with the picture that the outflowing ionized gas sourcing the broad nebular emission have more ordinary conditions; most likely $\nele$ is lower there, and the effect of H Balmer optical depth is insignificant due to an open geometry. It is unclear whether a broad component is important for the major UV emission lines such as {\rm C III]}$\lambda\lambda$1906,1908, {\rm N III]}$\lambda$1750 and {\rm [O III]}$\lambda\lambda$1660,1666. The SNRs in the MUSE data are not as high as in \cite{Choe2024}, but it does appear that FWHM$< 200$~km/s for these lines. It is likely the broad component is sub-dominant in the UV lines just like in many optical lines such as the auroral lines. Such a component may also be dusty and not subject to removal by UV radiation pressure (see Section \ref{sec:dust}), and should this component also have a fairly low overall covering, the geometry may allow for dust reddening of the broad lines while retaining a strong stellar UV continuum and nebular UV emission lines.

As described in Section \ref{sec:yeffect}, the H Balmer lines experience significant optical depth. The subsequent resonant scattering of the Balmer lines and Balmer self-absorption are expected to broaden the H Balmer lines to $100$'s~km/s, and may provide an explanation for the very broad component of the emission. Moreover, Balmer emission in the neutral gas resulting from fluorescence by the stellar continuum near higher-order H Lyman lines \citep[e.g.,][]{Luridiana2009} may also result in broad H Balmer lines, and this fluorescence is seen in our CLOUDY models. However, it is unclear if CLOUDY’s implementation of these effects is accurate and how broad these lines are expected to be. Furthermore, unknowns such as the column density or kinematics of the neutral region which could strongly affect this. Further detailed work is required to ascertain whether the presence of very broad $\sim500$~km/s line emission is quantitatively consistent with our physical picture.

In this work, we choose to neglect a possible nebula component coming from outflowing photo-ionization clouds that may be responsible for the broad emission line components reported in \cite{Choe2024}, with the justification that we have used in our analysis only the upper limits for the optical lines expect for {\rm [Ne III]}$\lambda$3869 and {\rm He I}$\lambda$5879. We leave a multi-nebular-component spectroscopic modeling covering from rest-frame UV to optical wavelengths, which is necessarily more sophisticated in terms of both data analysis and theoretical modeling, to future work when the JWST NIRSpec IFU data is incorporated into the analysis.

\subsection{Ly$\alpha$-pumped Fe III Fluorescence}
\label{sec:bowen}

\cite{Vanzella2020Tr} discovered remarkably bright fluorescent emission lines of Fe III from Godzilla, which are shown to result from Ly$\alpha$ excitation of a nearby Fe III$\lambda$1214 transition~\citep{Johansson2000FeIIIforbiddenlines}. To our knowledge, the only detailed report of this exact Fe III fluorescence process came from the Weigelt Blobs in $\eta$ Car's inner ejecta~\citep{Zethson2012Weigeltlines}, which are known to be heavily-CNO-processed material expelled by $\eta$ Car. Similar to what we have found for Godzilla's nebula, the fluorescent Weigelt Blobs host dense ionized gas $\nele\sim 10^{(7-8)}\,{\rm cm}^{-3}$~\citep{Hamann2012EtaCarInnerEjecta} and exhibit large enhancement in N and He abundances~\citep{Verner2005WeigeltBD}, and shows fluorescent O I lines by Ly$\beta$ pumping~\citep{Hamann2012EtaCarInnerEjecta}. Godzilla's nebula gas appears to physically and chemically resemble the Weigelt Blobs, but is excited by a much brighter radiation source. This similarity perhaps explains why both are conducive to Ly$\alpha$-pumped Bowen fluorescence.

Given the large $\sim 270\,{\rm km}\,{\rm s}^{-1}$ frequency de-tuning between H Ly$\alpha$ and the excited Fe III$\lambda$1214 line, a strong and broad Ly$\alpha$ line has to form within the nebula. This requires a geometry in which Ly$\alpha$ photons are spatially trapped by repeated resonant scattering after they are produced from recombination. This also requires a dust-free condition so that the Ly$\alpha$ photons are not internally destroyed. In the case of Godzilla, ionization-bounded hot bubbles embedded in the cluster gas, which we postulate as the formation sites of the nebular emission lines, may provide the right geometry as the surrounding self-shielded component of the cluster gas is optically thick to Ly$\alpha$ photons. The cluster gas is also required to be dust-free, which appears consistent with non-detection of significant reddening of Godzilla's FUV continuum. We therefore suggest, at least qualitatively, that our physical picture for Godzilla's nebula may explain the observed Ly$\alpha$-pumped Fe III fluorescence. We plan to investigate this in more details in a future publication.

\subsection{Removal of Dust Grains}
\label{sec:dust}

Our inferred relatively low ${\rm E(B-V)}<0.08$ implies that the overall nebula environment of Godzilla is not extremely dusty and is largely transparent to FUV. A grain-free condition is also required in the active zones to allow for H Lyman-series pumping and fluorescence.

As our photoionization modeling implies, direct stellar irradiation, dominated by FUV in energy, may be intense enough to vaporize dust grains. For the high radiation intensity, we neglect the effects of time-dependent temperature spikes for small grains, and only consider an equilibrium-state grain temperature $T_d$, which can be found by balancing the rate of FUV heating and the rate of cooling by thermal re-radiation at infrared wavelengths~\citep{Draine2011ISMtext}:
\begin{align}
    4\pi\,a^2\,\langle Q \rangle_{T_d}\,\sigma_{{\rm SB}}\,T^4_d = \pi\,a^2\,\langle Q \rangle_{\rm UV}\,F_{\rm UV}.
\end{align}
Here $\langle Q \rangle_{T_d}$ the absorption efficiency averaged over a Planck spectrum of temperature $T_d$, and $\langle Q \rangle_{\rm UV}$ is the absorption efficiency averaged over the incident radiation, whose flux is dominated by UV photons, and $\sigma_{{\rm SB}}$ is the Stefan-Boltzmann constant.

The precise values of $\langle Q \rangle_{T_d}$ and $\langle Q \rangle_{\rm UV}$ depend on grain size, grain composition, and the spectral shape of the incident star light, and are not always monotonic. For FUV photons $\langle Q \rangle_{\rm UV} \simeq 1$. According to Figure 24.3 of \cite{Draine2011ISMtext}, we conservatively estimate:
\begin{align}
    \langle Q \rangle_{T_d} < 0.5\,\left(\frac{a}{\mu{\rm m}}\right),
\end{align}
in the temperature range $100< T_d/{\rm K} < 1000$, for sufficiently large grains $a>0.01\,\mu{\rm m}$, and for both silicate and carbonaceous grains. These conservative limit leads to
\begin{align}
    & T_d > \left( \frac{(\mu{\rm m}/a)}{0.5}\,\frac{\chi\,\Phi({\rm H}^0)\,\langle h\nu \rangle_{\rm ion}}{4\,\sigma_{\rm SB}}\right)^{\frac14} \nonumber\\
    & = 1200\,{\rm K}\, \left(\frac{a}{0.01\,\mu{\rm m}}\right)^{-\frac14}\nonumber\\
    & \times \left(\frac{\chi}{10}\right)^{\frac14} \left(\frac{\Phi({\rm H}^0)}{10^{15.9}\,{\rm s}^{-1}{\rm cm}^{-3}}\right)^{\frac14} \left(\frac{\langle h\nu \rangle_{\rm ion}}{20\,{\rm eV}}\right)^{\frac14}.
\end{align}
Here we have introduced the dimensionless $\chi$, the ratio between the FUV flux and the ionizing flux. The lower bound is $T_d>680\,$K for $a=0.1\,\mu{\rm m}$ and $T_d>380\,$K for $a=1\,\mu{\rm m}$. Given the sublimation temperature $\sim 1200\,$K for silicates and $\sim 1800\,$K for carbonaceous grains, small silicates grains with $a \lesssim 0.01\,\mu{\rm m}$ could have been all evaporated by radiation, but large silicate grains or carbonaceous grains can likely survive.

For the geometry of embedded HII bubbles (\refsec{core}), small silicate grains $a\lesssim 0.01\,\mu$m should be vaporized by individual O stars in the HII bubble vicinity, but it is unclear if the global FUV background throughout the retained gas cloud is intense enough to vaporize the other grains. Assuming that a fraction $x_{\rm FUV}$ of the FUV sources are enclosed within radius $R_g$, the mean FUV irradiation is a fraction $\simeq 0.11\,(x_{\rm FUV}/0.1)\,(R_g/0.2\,{\rm pc})$ of the FUV irradiation in individual HII bubbles.

It may be that over time radiation pressure pushes all surviving grains out of the gas cloud. For grain collision with hydrogen atoms in an HI gas, \cite{DraineSalpeter1979GrainsInHotGas} introduced the function
\begin{align}
    G(s) = \frac{8\,s}{3\,\sqrt{\pi}}\left(1 + \frac{9\pi}{64}\,s^2\right)^{1/2}.
\end{align}
Here $s=v_d/\sqrt{2\,\kB\,T/m_p}$, where $v_d$ is the grain drift velocity relative to the gas and $T$ is the gas temperature. When the drift velocity is supersonic $s\gg 1$, $G(s)=s^2$. Following \cite{Draine2011HIIRegionRadiationPressure}, we find a drift velocity
\begin{align}
    & v_d = 22\,{\rm km}\,{\rm s}^{-1}\,\left(\frac{\chi}{10}\right)^{\frac12}\,\left(\frac{\Phi({\rm H}^0)}{10^{14.9}\,{\rm s}^{-1}\,{\rm cm}^{-2}}\right)^{\frac12}\,\left(\frac{\langle h\nu \rangle_{\rm ion}}{20\,{\rm eV}}\right)^{\frac12} \nonumber\\
    & \times \left(\frac{\langle Q_{\rm pr} \rangle}{1.5}\right)^{\frac12}\, \left(\frac{n_V}{10^{6.5}\,{\rm cm}^{-3}}\right)^{-\frac12},
\end{align}
where $\langle Q_{\rm pr} \rangle$ is the wavelength-averaged absorption efficiency. This is not small compared to the turbulence velocity $u$, so let us assume that turbulence motion does not significantly prohibit grain ejection. After a short time $R_g/v_d \approx 10^4\,{\rm yr}\,(R_g/0.2\,{\rm pc})/(v_d/20\,{\rm km}\,{\rm s}^{-1})$, FUV radiation should eject all grains out of the retained dense gas if they have survived sublimation. 

In this process, dust grains drift relative to the gas at terminal speeds and exert an outward force on the gas as well. However, gravity of the cluster gas and the stars it encloses is strong enough to prevent gas from being ejected together with the dust grains. Indeed, the characteristic inward pressure due to gravity can be crudely estimated as
\begin{align}
    & P_{\rm grav} \gtrsim \frac{G\,M^2_g}{R^4_g} \nonumber\\
    & \simeq 1.5\times 10^{14}\,{\rm K}\,{\rm cm}^{-3}\,\left(\frac{M_g}{10^5\,M_\odot}\right)^2\,\left(\frac{R_g}{0.2\,{\rm pc}}\right)^{-4}.
\end{align}
In comparison, the typical outward pressure from UV radiation transferred through the grain-gas coupling is significantly smaller
\begin{align}
    P_{\rm UV} \simeq 10^{11}\,{\rm K}\,{\rm cm}^{-3}\,\left(\frac{\Phi({\rm H}^0)}{10^{15.1}\,{\rm s}^{-1}\,{\rm cm}^{-2}}\right)\,\left(\frac{\langle h\nu \rangle_{\rm ion}}{20\,{\rm eV}}\right)\,\left(\frac{\chi}{10}\right).
\end{align}
 

\subsection{Godzilla and Multiple Stellar Populations}

The high mass and compactness of Godzilla make it a likely candidate for an analog progenitor of globular clusters~\citep{Kruijssen2014GCformation}, which ubiquitously host multiple stellar populations of distinct light element abundances (see \cite{Bastian2018ARAAGCMultiPop} for a review).
The dense nebula gas of Godzilla likely has accumulated inside the cluster, even at an age $t_{\rm age}=4$--$6\,$Myr when the most massive stars have already started CCSNe. While the standard expectation is that CCSNe will quickly clear out any cluster gas~\citep[e.g.,][]{Calura2015WindFeedback, Dercole2010}, the successful retention of stellar ejecta in Godzilla, if confirmed, will corroborate theoretical predictions that in dense star clusters feedback sometimes fails in doing this job \citep[e.g.,][]{Krause2013,Tenrio2016}, leaving behind a gas reservoir conducive to forming new generation of stars.

However, it does not seem likely to us that this particular dense nebula gas we are witnessing in Godzilla will birth the typical second generation (2P) stars. Assuming that CCSNe fail to expel the cluster gas, as appears to be the case in Godzilla, the strong turbulence driven by CCSNe feedback is expected to prohibit gravitational collapse \citep{Hayward2016}, disfavoring formation of 2P stars after CCSNe onset. While some propose that this problem can be circumvented in the first $5$--$10\,$Myr of the cluster evolution by delaying the onset of CCSNe \citep[e.g.,][]{Renzini2022}, our inferred cluster age suggests that the delay of CCSNe onset may either be at most $1$--$2\,$Myr or does not happen at SMC-like stellar metallicities. Even if star formation still proves possible after CCSN onset, the observed simultaneous N and O enrichment is not compatible with the abundance pattern of the typical 2P stars, as they typically show N enhancement but no enhancement in $\alpha$-elements. Should Godzilla be a true globular cluster progenitor, 2P star formation must have already occurred prior to the onset of CCSNe.

The observed enhancements in N/O and He/H should be primarily driven by wind ejecta (via comparison to \citetalias{Molla2012StarClusterEjecta}, see Figure~\ref{fig:molla}). While our inferred N/O is broadly consistent with the 2P stars in globulars (Fig.~\ref{fig:all_abuns}), the inferred He/H value is likely too high. The He abundance in globulars typically has a small spread $\Delta Y \lesssim 0.1$, implying less than unity enhancement in the He/H ratio. By contrast, we infer a factor of $\sim 2$--$4$ enhancement of He/H in Godzilla. Intriguingly, in some of the more special globular clusters such as $\omega$~Centauri, He/H differs by as much as a factor of two between multiple populations \citep{Dupree2013}, although it is unclear whether Godzilla could be a progenitor of such unique GCs. It is possible that the observed dense nebular gas is unrelated to 2P star formation, and the typical 2P star formation occurs even earlier, prior to the onset of He-rich winds from evolved massive stars. The models of \citetalias{Molla2012StarClusterEjecta} find this onset at $\sim2.5\,$Myr, at which point CNO-processed material is ejected and enriches the surrounding with N and He. However, we note that there are likely earlier sources unaccounted for in the \citetalias{Molla2012StarClusterEjecta} models that produce ejecta rich in N but not in He, such as Very Massive Stars \citep{Vink2023} or supermassive stars \citep{Charbonnel2023gnz11}. These sources would allow for similar N/O while significantly reduced (but still mildly enhanced) He/H to match what is observed in the majority of the GCs.

\section{What Is the Magnification of Godzilla?}
\label{sec:mag}

The magnification factor of Godzilla is a highly contested topic. Precise knowledge of or empirical constraints on its value will have strong implications for the physical nature of this object. Even though this work focuses on the spectroscopic interpretation of Godzilla, we feel compelled to discuss below how the magnification factor may be empirically constrained.

\subsection{Constraints From Flux Ratios}
\label{sec:ratio}

\begin{figure*}[t]
    \centering
    \includegraphics[width=18cm]{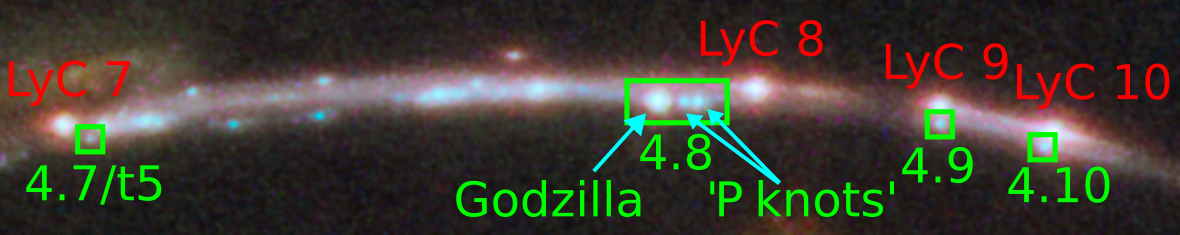}
     \caption{HST false color image of Godzilla, the `P knots', and candidate counterimages identified by \cite{Sharon2022SunburstLensModel} and \cite{Diego2022godzilla}; the labeling reflects the the naming scheme of both papers. Godzilla is significantly brighter than its counterimages, implying it is highly magnified by comparison. The color of the counterimages may more closely match the color of the P knots, which may indicate that Godzilla is intrinsically fainter than the P knots and is blended with and dominated by them in the counterimages. This figure is modeled after \cite{Sharon2022SunburstLensModel}'s Figure 2 and \cite{Choe2024}'s Figure 1, which both show color images using different HST or JWST filters, and provide useful further reference for the relative colors of Godzilla, the P knots, and the candidate counterimages.}
    \label{fig:godzilla_counter}
\end{figure*}

Due to the anomalous lensing configuration required, general cluster-scale lensing models such as those in \cite{Pignataro2021SunburstLensModel} and \cite{Sharon2022SunburstLensModel} do not give trustworthy predictions of the magnification factor for Godzilla. However, these models can leverage candidate counter-images of Godzilla to infer a magnification on the basis of the observed flux ratio between Godzilla and a possible counter-image, as performed in \cite{Diego2022godzilla}. Simply put, assuming the macro lens model for a counter-image is reliable (because the counter-image typically is less magnified and further from critical curves), the flux ratio between Godzilla and its counter-image can be used to infer the magnification of Godzilla. Following this approach, \cite{Diego2022godzilla} find a minimum magnification of Godzilla $\mu \sim 600$ pulling from the lens model of \cite{Pignataro2021SunburstLensModel}, and an even higher $\mu \sim 1400$ or $\mu \sim 8000$ using the \texttt{WSLAP+} lens modeling approach depending on the sets of constraints used. We note that this work effectively measured the magnification from the counterimage with the lowest fractional flux errors, and that the median across the counterimages typically skewed lower (see their Figure 4). We also repeat exactly the approach of \cite{Diego2022godzilla} with the lens model predictions of \cite{Sharon2022SunburstLensModel}, which uses the same modeling software as \cite{Pignataro2021SunburstLensModel}, to infer a lower minimum magnification $\mu \sim 300$. 

As is apparent, magnification predictions of these published Sunburst lens models vary by greater than an order of magnitude. This may result from families of degeneracies stemming from the mass-sheet degeneracy \citep{Falco1985, Oguri2003} , which hinder absolute magnification estimates for lens systems without a high number of multiply-imaged sources at a number of redshifts, such as PSZ1 G311.65-18.48 which lenses the Sunburst galaxy. It may also be subject to other lensing systematics, such as milli-lensing by optically invisible dark matter subhalos \citep{Dai2020S1226millilens} or line of sight structure \citep{Dalal2005} which are generally unaccounted for in strong lens models. With the above caveats in mind, we caution the reader for the following flux ratio estimates.

The candidate counter-images of Godzilla are not yet confirmed, however it is suggested that in the counter-images the source object of Godzilla may be blended with two neighboring knots, dubbed the ``P knots'' in \cite{Diego2022godzilla}, which themselves are likely a lensed image pair straddling a critical curve. \cite{Diego2022godzilla} propose 5 likely counter-images (see their Fig. 3) labeled t1-t5, while \cite{Sharon2022SunburstLensModel} propose a different set of 9 counterimages (their image system 4, see their Figure 2) named 4.1-4.10, where 4.8 corresponds to Godzilla. The two sets share only one candidate counter-image, which corresponds to t5 in \cite{Diego2022godzilla} and 4.7 in \cite{Sharon2022SunburstLensModel}. We opt to adopt the counter-image choice and nomenclature of \cite{Sharon2022SunburstLensModel}, which are shown in Figure \ref{fig:godzilla_counter}, as this seems more consistent with the HST and JWST colors \citep[see][]{Choe2024}.

Adopting these images, we follow the magnification ratio argument of \cite{Diego2022godzilla}, and restrict ourselves to 4.7, 4.9, and 4.10 for simplicity. There are three lens models from which the magnification of these sources can be inferred, however \cite{Diego2022godzilla} only makes a prediction for 4.7 (corresponding to t5). Using the PSF photometry approach described in Sec. \ref{sec:phot}, we measure the flux ratio of Godzilla, the P knots, 4.7, 4.9, and 4.10 in the F814W filter, which are shown in Table \ref{tab:tab_models}. We use these flux ratios to compute the Godzilla's magnification, via 4.7, 4.9 and 4.10 separately, as the use of each candidate counter-image may be affected by systematics differently. The flux ratio of 4.7 to Godzilla is $\sim 2\%$ (which was similarly found by \cite{Diego2022godzilla}), and the magnification of 4.7 ranges from $\mu_{4.7}=5.2^{+1.0}_{-0.4}$ from \cite{Sharon2022SunburstLensModel} to $\mu_{4.7} \sim 105$ from \cite{Diego2022godzilla}. This would in turn imply a lower limit on Godzilla's magnification $\mu \gtrsim 250$--$5000$ as Godzilla may not account for the totality of the flux of 4.7. Similarly for 4.9, which has $\sim 8\%$ the flux of Godzilla, magnification predictions are $\mu_{4.9} = 31^{+9}_{-3}$ from \cite{Sharon2022SunburstLensModel} and $\mu_{4.9} = 71.1^{+7.6}_{-6.4}$ from \cite{Pignataro2021SunburstLensModel}, while \cite{Diego2022godzilla} do not make a prediction for this image. Here, we require $\mu \gtrsim 390-890$. Finally we evaluate 4.10, which has $18\%$ the flux of Godzilla (more than twice the flux of 4.9), despite the magnification predictions of \cite{Pignataro2021SunburstLensModel} and \cite{Sharon2022SunburstLensModel} being similar to those for 4.9. This yields $\mu_{4.10}=34^{+10}_{-3}$ and $\mu_{4.10} = 86.6^{+7.6}_{-6.1}$ respectively. This then implies a magnification for Godzilla $\mu \gtrsim 190-480$. These results are summarized in \reftab{tab_models}. 

Lens model uncertainties vastly dominate this approach as apparent from the comparison between different lens models. While it is highly probable that Godzilla is magnified by factor of hundreds or greater, it is far from obvious to us that Godzilla must be magnified by thousands or even upwards of ten thousand. A more intermediate magnification of $\mu=1000$ is not only consistent with the measured flux ratios as we have discussed, but would also be high enough to cause the source of the P knots to dominate the color of the counter-image. Indeed, the source of P knots is likely a more massive star cluster than Godzilla. This is corroborated by the ratio of the 4.9 and 4.10 fluxes to the neighboring two lensed images of the LyC star cluster, which we will refer to as LyC 9 and LyC 10 respectively. Given that 4.9 and 4.10 lie radially to LyC 9 and LyC 10, it is expected that each pair likely have similar magnification factors. This statement is independent of the lens model uncertainties cautioned above, and indeed, both \cite{Pignataro2021SunburstLensModel} and \cite{Sharon2022SunburstLensModel} predict this to be the case (despite disagreeing significantly on the magnification value itself). \cite{Pascale2023} and \cite{Vanzella2022SunburstEfficiency} both find that the LyC cluster is likely extremely massive $\gtrsim 10^7 M_\odot$. We find that 4.9 and 4.10 are each $\sim 20\%$ the flux of their neighboring lensed image of the LyC star cluster, implying that 4.9 and 4.10 must host $\gtrsim 10^6 M_\odot$ in stellar mass. We find it plausible that, due to the complex caustic structure, the P knots are much less magnified than Godzilla (indeed they are similar in brightness to within a factor of 2 to 4.9 and 4.10, which are not unusually magnified), yet their underlying source is a more massive star cluster which outshines Godzilla in the candidate counter-image 4.7.

\begin{deluxetable}{cccccc}
\tabletypesize{\footnotesize}
\tablecaption{Magnifications and Flux Ratios of Godzilla Counterimages}
\label{tab:tab_models}
\tablecolumns{6}
\tablehead{
\colhead{\bf ID} &   \colhead{{\bf $\mu_{\rm Sharon}^{a}$}} & \colhead{\bf $\mu_{\rm Pignataro}^{b}$} & \colhead{\bf $\mu_{\rm Diego}^{c}$} & \colhead{\bf $f^{d}/f_{\rm Godzilla}$ } & \colhead{\bf $\mu_{\rm Godzilla}$}} 
\startdata
4.7/t5 & $5.2^{+0.2}_{-0.9}$ & $[11,19]^{e}$ & $19, 105^{f}$ & $0.02$ & $[250,5000]$ \\
4.9 & $31.1^{+9.9}_{-13.1}$ & $71.1^{+7.6}_{-6.4}$ & -- & $0.08$ & $[390,890]$ \\
4.10 & $33.9^{+10.1}_{-12.9}$ & $86.6^{+7.6}_{-6.1}$ & -- & $0.18$ & $[190,480]$ \\
\enddata
\tablecomments{The ID's (following \citealt{Sharon2022SunburstLensModel}), lens model predicted magnifications, and measured flux ratios with respect to Godzilla for the candidate Godzilla counterimages. We also list the range of magnification lower limits for Godzilla implied by the lens model magnifications and measured flux ratios (see Section \ref{sec:ratio}). If Godzilla's counter image is blended with the counter image of the P knots and contributes to less than a fraction $f$ of the flux, then the implied value for $\mu_{\rm Godzilla}$ should be increased by a factor $1/f$.}
\tablenotetext{a,b,c}{$\quad \quad$\cite{Sharon2022SunburstLensModel}, \cite{Pignataro2021SunburstLensModel}, and \cite{Diego2022godzilla} respectively.}
\tablenotetext{d}{Measured in HST ACS/WFC F814W. Photometric uncertainties are generally $\lesssim 10\%$ and lens model uncertainty dominates the magnification estimate.}
\tablenotetext{e}{Magnification not given for 4.7, we instead cite their 5.1g and 5.2g which bracket 4.7 as the lower and upper limits respectively.}
\tablenotetext{f}{Magnifications given for the two different lens models M1 and M2 in \cite{Diego2022godzilla}.}
\end{deluxetable}

\subsection{Constraints From Flux Variability}

In the high magnification regime, the exact value of Godzilla's magnification and its age has strong implications for the expected flux variability on the order of days to weeks due to microlensing by intracluster stars~\citep{Venumadhav2017CausticMicrolensing, Diego2018DMUnderMicroscope, Oguri2018CausticMicrolensing}. These intracluster stars cast a fine network of micro caustics which cause a spatial pattern of magnification on the source plane that is highly non-uniform. As the entire star cluster moves across this pattern, each cluster member star will acquire a time-varying magnification factor, and such magnification factors behave uncorrelated between different member stars~\citep{Dai2020S1226millilens}. Then, what can be measured is the superposition of variable fluxes from all stars in the star cluster. \cite{Dai2021Sunburst} develops a quantitative framework to describe the statistics of this collective variability, and concludes that when observed at many random times the total flux has a standard deviation that is proportional to the standard deviation of the relative magnification factor fluctuation $\delta=\mu/\bar{\mu}-1$ and is inversely proportional to the square root of the number of stars. At fixed observed magnitude, a more magnified star cluster must have less stars, which then implies larger fractional fluctuations in the flux. On the other hand, a higher macro magnification factors cause the micro caustics to crowd together, which increases the standard deviation of $\delta$.

\begin{figure}[ht]
    \centering
    \includegraphics[width=\columnwidth]{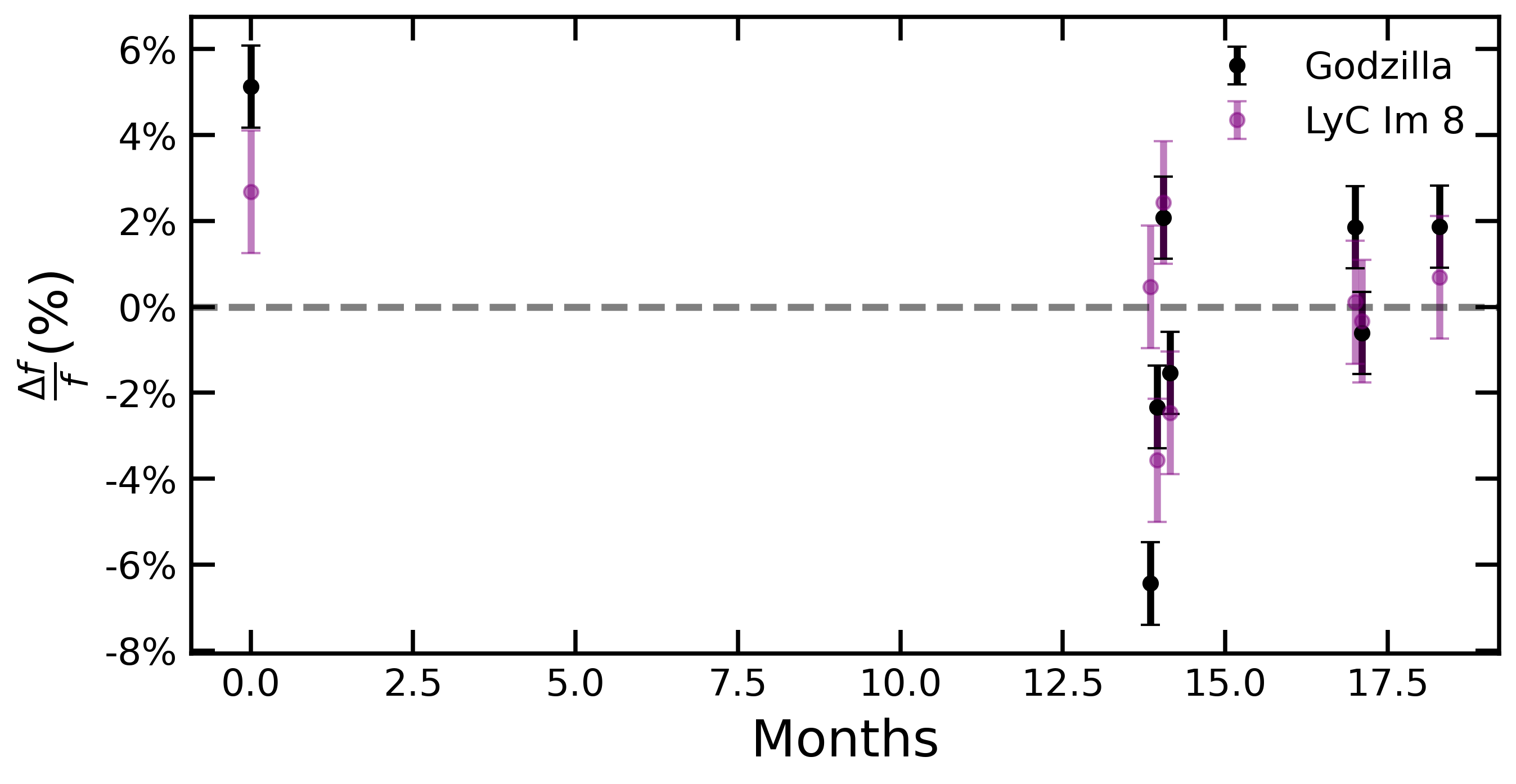}
    \caption{Fractional flux changes with respect to the mean for Godzilla (black) and image 8 of the LyC cluster (purple) sampled across 8 visits of HST WFC3/IR F140W. These 8 visits make up 4 distinct epochs relative to the timescales for microlensing fluctuations ($\gtrsim$ few days), with the first visit occurring on June 24, 2019. We observe that image 8 of LyC, which is not expected to showcase significant microlensing-induced variability, fluctuates at the $\sim 2\%$ level, while Godzilla fluctuates marginally greater at the $\sim 3\%$ level and shows the largest fractional change between any two visits of $\sim 13\%$. We note that the visit with the greatest factional change of $\sim 8\%$ at 14 months is accompanied by 3 other visits with comparatively smaller fractional changes. These 4 visits take place over the span of 2 days, a timescale which is likely too short for microlensing to induce significant variability.}
    \label{fig:variable_phot}
\end{figure}

\cite{Sharon2022SunburstLensModel} pointed out that Godzilla shows no signs of significant variability in multiple HST visits over a 2-year period. While it is clear from that work that Godzilla does not drastically brighten or fade, the next question is whether it exhibits subtle flux variability in the continuum. We seek to quantify this by making use of 8 imaging visits in the HST WFC3\/IR F140W filter, which sample $\sim 4$ distinct time frames (i.e., separated by at least a day) across a time window of 1.5 yr. The 8 visit dates are as follows: 06/24/2019, 08/29/2020, 08/30/2020, 8/30/2020, 8/31/2020, 11/11/2020, 11/15/2020, and 12/30/2020. We align the image of each visit following the methodology outlined in Sec.\ref{sec:data}, but to the native WFC3/IR pixel scale of 0.13\arcsec. We measure photometry for Godzilla within a fixed aperture of radius 0.4\arcsec, and estimate the sky background in each visit placing the same aperture on a dozen locations off the arc in the blank sky. To estimate the fractional flux change from visit to visit, the mean flux across the visits is then set to the measured F140W flux from PSF photometry, as the aperture fluxes contain arc background which may otherwise dilute the fractional change. As a control sample, we repeat this procedure on the nearby image 8 of the LyC star cluster (LyC 8 hereafter) for an estimate of photometric systematics (e.g. PSF variation, subpixel misalignments) which certainly dominate over statistical errors. This is justified because the lower magnification of LyC 8 but higher number of stars of the LyC star cluster together imply microlensing-induced flux variations at the percent level or less \citep[see discussion in][]{Dai2021Sunburst}. The fractional flux variations are shown in Figure \ref{fig:variable_phot}. We infer systematics at $2\%$ from LyC 8. Intriguingly, Godzilla exhibits fluctuations marginally greater than this systematic uncertainty with a standard deviation of $\sim 3\%$ and a maximum range $\sim 13\%$. Figure \ref{fig:variable_phot} shows that the majority of the fluctuations between Godzilla and LyC 8 are correlated, this may be due to a flux calibration systematic or incomplete background estimation, but indicates the true flux variations may very well be closer to the $1\%$ level.

To constrain Godzilla's magnification by assuming that microlensing must induce flux variations with a standard deviation $<3\%$, we need to estimate the surface density of intracluster stars by measuring the surface brightness of the intracluster light (ICL) \citep[e.g.,][]{Kelly2018Icarus, Meena2023Flashlights}. \cite{Dai2021Sunburst} estimated the ICL stellar population from MUSE IFU spectroscopy near the brightest cluster galaxy (BCG) and found a stellar age $\sim 5$~Gyr and metallicity $Z \sim 0.4\,Z_{\odot}$. However, at the distance of the Sunburst arc ($\sim 30 \arcsec$), \cite{Dai2021Sunburst} judged that the ICL may fall below the diffuse sky background ($\sim25$mag/arsec$^2$ in F160W), resulting in an upper limit on the averaged convergence $\kappa_{\star} < 0.002$ from intracluster stars.

\begin{figure}[ht]
    \centering
    \includegraphics[width=\columnwidth]{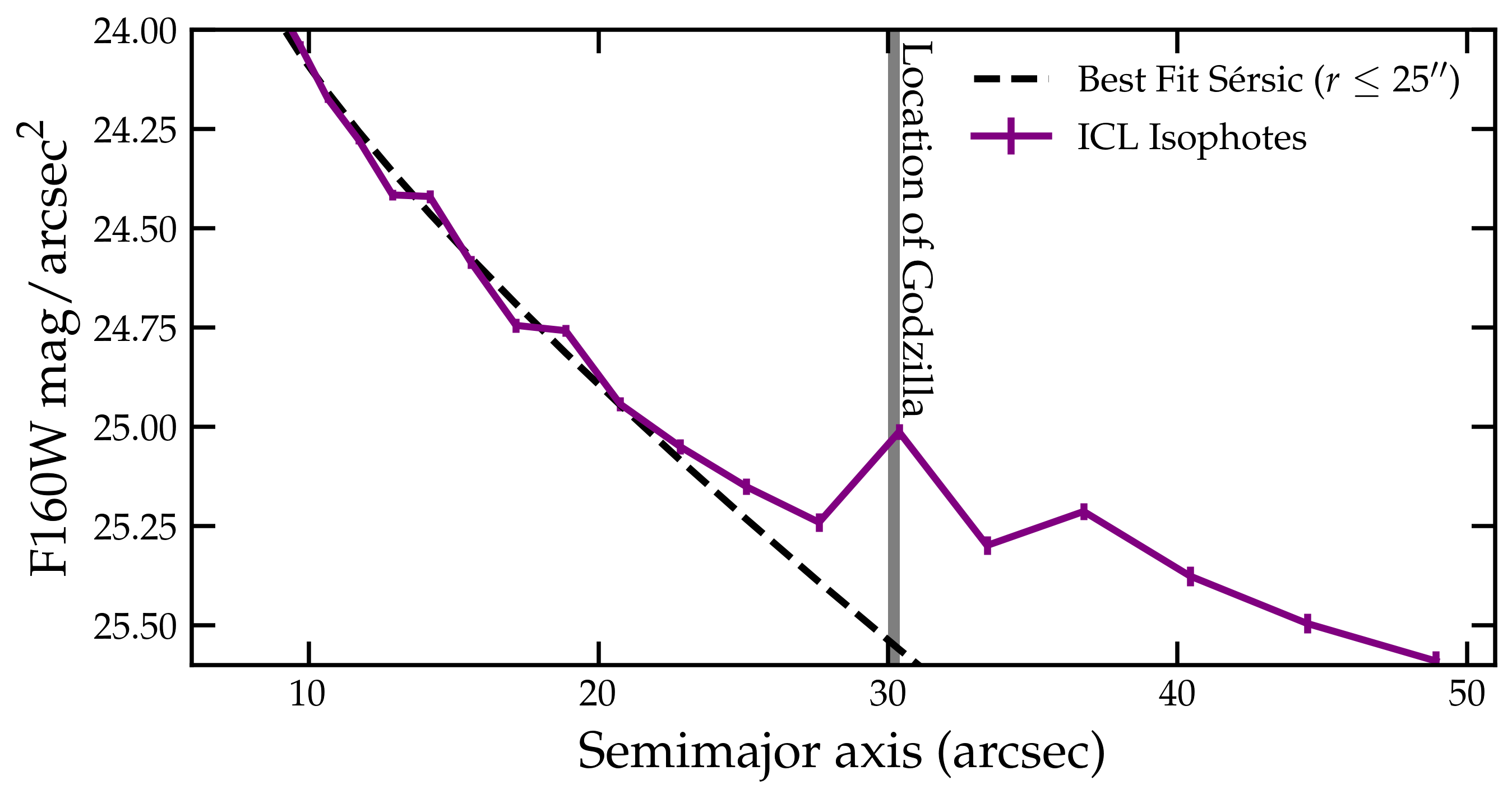}
    \caption{Intracluster light (ICL) surface brightness profile in HST WFC3/IR F160W measured via elliptical isophotes (purple) centered at the brightest cluster galaxy (BCG). The surface brightness profile begins to flatten at a semimajor axis $\sim25\arcsec$, which may indicate the ICL is approaching the diffuse sky background. To conservatively estimate the ICL contribution at the location of Godzilla ($\sim 30\arcsec$), we fit a S\'{e}rsic profile to isophotes with semimajor axis $a \leq 25\arcsec$ and extrapolate to the location of Godzilla. We find an ICL surface brightness $\sim 25.5{\rm ~mag/arcsec}^2$, corresponding to an average convergence $\kappa_{\star} \gtrsim 0.001$ from intracluster stars. We note the surface brightness profile shows a peak at the location of the Sunburst arc due to the diffuse arc background not well masked by SExtractor, where the arc generally makes up a large fraction of the isophote at that semimajor axis.}
    \label{fig:icl}
\end{figure}

To improve, we fit elliptical isophotes to the ICL in F160W imaging following the methods of \cite{Montes2021}. Briefly, we run \texttt{Source Extractor} \citep{Bertin1996} following a \texttt{HOT+COLD} approach \citep[e.g.,][]{Merlin2016, Pascale2022} to mask bright sources which may contaminate the fit. Stars are identified using the results of \texttt{DAOStarFinder} from Sec. \ref{sec:phot} and their masks are vetted by eye to be sufficiently large; features such as the PSF wings and diffraction spikes are typically not well masked by \texttt{Source Extractor} and are further masked by hand. Elliptical isophotes are then fit to the masked image using the \texttt{photutils.isophote} package, which implements the algorithm of \cite{Jedrzejewski1987}. As seen in Figure \ref{fig:icl}, we find that the ICL at a distance of $\sim 25\arcsec$ from the BCG the ICL profile begins to flatten. It is not unusual for the ICL to be composed of both a steep and shallow components, as shown in \cite{Montes2021}, however this may instead indicate it is approaching the diffuse sky background. To be conservative, we fit a S\'{e}rsic profile \citep{Sersic1968} to the measured surface brightness from isophotes only up to $\sim 25\arcsec$. From this, we extrapolate to the location of Godzilla and infer an underlying ICL surface brightness $25.5$~mag/arcsec$^2$, implying $\kappa_{\star} \gtrsim 0.001$. For the following tests, we assign $\kappa_{\star} = 0.001$ as a conservative lower limit, emphasizing that a larger $\kappa_{\star}$ would induce larger variability and hence imply a decreased upper limit on Godzilla's magnification.

The timescale of variability is set by the effective source velocity (see Eq.(12) of \cite{Venumadhav2017CausticMicrolensing}, which however does not account for the observer's motion)
\begin{align}
\label{eq:vt}
    v_t = \left| \left( \frac{\bfvs}{1+z_s} - \frac{\Ds}{\Dl}\,\frac{\bfvl}{1+z_l} + \frac{\Dls}{\Dl}\,\frac{\bfvo}{1+z_l} \right) \cdot \hat{\boldsymbol{s}} \right|.
\end{align}
Here $\Dl$, $\Ds$ and $\Dls$ are angular diameter distances to the lens, to the source, and from the lens to the source, respectively. $\bfvs$, $\bfvl$ and $\bfvo$ are peculiar velocities of the source, the lensing galaxy cluster, and the Solar System, respectively. The unit vector $\hat{\boldsymbol{s}}$ is the direction perpendicular to the majority of the highly stretched micro caustics on the source plane. While the value of $v_t$ for Godzilla is not known, the probability distributions for $\bfvs$ and $\bfvl$ can be predicted from the theory of large-scale structure formation assuming Planck 2015 cosmology~\citep{Planck2015cosmology}. Furthermore, we infer $\bfvo$ from the measurement of the CMB temperature dipole~\citep{Saha2021LocalMotionPlanck2018}. We find that $v_t$ is drawn from a normal distribution with a standard deviation $\sigma_{v_t}\approx 600\,{\rm km}/{\rm s}$, but the sign of $v_t$ is irrelevant.

\begin{figure}[ht]
    \centering
    \includegraphics[width=\columnwidth]{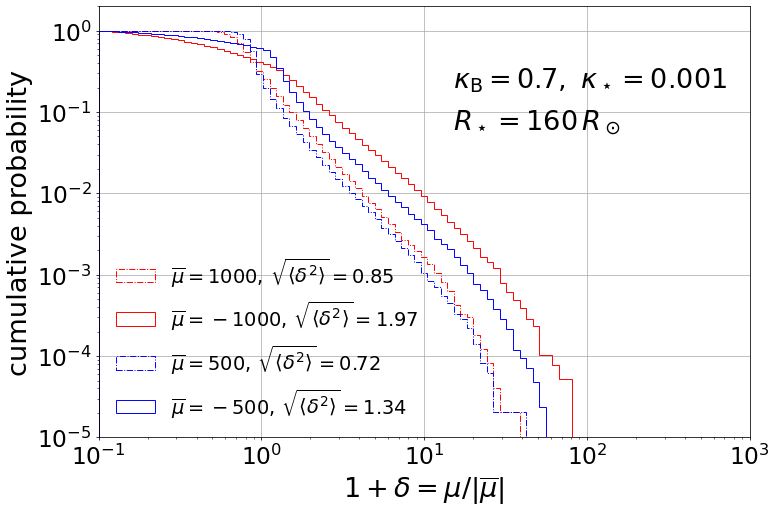}
    \caption{Probability distributions of $1+\delta=\mu/\bar{\mu}$ for any individual source star, derived separately for a pair of macro images with macro magnification factors $\bar\mu=\pm 500$ (red) and $\bar\mu=\pm 1000$ (blue). The value of $\sqrt{\langle\delta^2\rangle}$ for each macro image is quoted in the legend text. We set a local macro convergence $\kappa_{\rm B}=0.7$, $\kappa_\star=0.001$, and use a relatively large average value for the stellar photosphere radius $R_\star=160\,R_\odot$ in order to be conservative in estimating $\sqrt{\langle\delta^2\rangle}$.}
    \label{fig:delta_stats_star_cluster}
\end{figure}

We run microlensing simulations following the methods of \cite{Dai2021Sunburst} to assess the probability of observing $\leq3\%$ flux variations given the cadence of the F140W imaging visits. As an example, Figure \ref{fig:delta_stats_star_cluster} shows the statistics of the random magnification factor induced by intracluster microlensing for any individual cluster member star, where Godzilla is an unresolved pair of images of a lensed star cluster with a total magnification factor $\mu = 2000$. In this case, $\sqrt{\langle\delta^2\rangle}=1.97$ ($0.85$) for the macro image of negative (positive) parity. Combining both images, we have a reduced effective $\sqrt{\langle\delta^2\rangle} = 1.5$. Our spectroscopic analysis suggests that Godzilla is a star cluster with $\log(\mu\,M_\star/M_\odot)=9.3$, for which we estimate from Figure 1 of \cite{Dai2021Sunburst} $\epsilon_2 \approx 0.03\,(\mu/2000)^{1/2}$ (using the notation therein), a factor depending on the luminosity function of stars. The standard deviation of the fractional flux variability is $\epsilon_2\,\sqrt{\langle\delta^2\rangle}=4.5\%$ for $\mu=2000$, which is a little larger compared to the measured standard deviation $<3\%$ we derive from data. If the total magnification of the image pair is $\mu=1000$, we find $\sqrt{\langle\delta^2\rangle} = 1.1$ and $\epsilon_2\,\sqrt{\langle\delta^2\rangle}=2.3\%$. The time intervals between imaging epochs are sufficiently long because in the lensed star cluster scenario the variability is rapid due to the large number of member stars, and hence the knowledge of $v_t$ is not required here. We therefore conclude that if Godzilla is a young star cluster its magnification factor is unlikely to be higher than $\mu=2000$.

\begin{figure}[ht]
    \centering
    \includegraphics[width=\columnwidth]{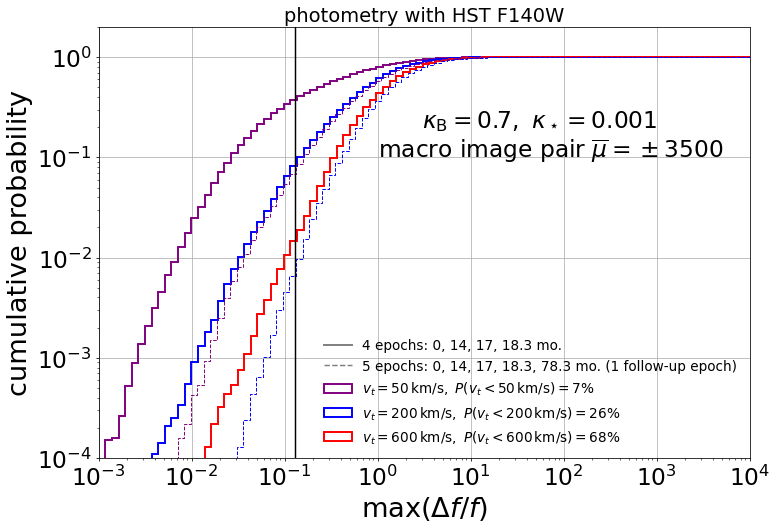}
    \caption{Probability distribution for the maximal observed flux variation expected for a single lensed star, derived from simulating the cadence pattern of the actual 4 distinct imaging epochs in F140W (solid curves). We set a local macro convergence $\kappa_{\rm B}=0.7$, $\kappa_\star=0.001$, and assume an unresolved macro image pair of $\mu_{\rm B}=\pm 3500$. The black vertical line marks the maximal fractional flux change $13\%$ between two epochs measured through aperture photometry. The results depend on the unknown effective velocity of the source $v_t$, which, in Planck 2015 cosmology, has a standard deviation $600\,{\rm km}\,{\rm s}^{-1}$ for the Sunburst system. The single lensed star scenario is compatible with data only if the effective velocity is unusually small e.g. $v_t = 50\,{\rm km}\,{\rm s}^{-1}$, and would be tightly constrained with another measurement in 2024 or later (dashed curves).}
    \label{fig:flux_change_cdf_single_star}
\end{figure}

We also consider the scenario that Godzilla is a single hyperluminous star very close to a caustic and the lensed image pair add up to a very high total magnification $\mu = 7000$, following the suggestion of \cite{Diego2022godzilla}. In Figure \ref{fig:flux_change_cdf_single_star}, we simulate the 4 distinct epochs corresponding to the HST F140W imaging data. For $v_t > 200\,{\rm km}/{\rm s}$ there is $<1\%$ chance of not seeing any flux change $>13\%$ among the F140W imaging visits. The chance increases to $>40\%$ for $v_t<50\,{\rm km}/{\rm s}$; however, we emphasize that this velocity range is quite unlikely, with only a $7\%$ occurrence chance a priori.

Figure \ref{fig:flux_change_cdf_single_star} also shows that having one more imaging visit in late 2024 or later will provide a long time baseline and will tightly constrain the scenario of a single lensed star even for very low $v_t$. Low values of $v_t$ are unlikely but certainly possible, so we suggest that the search for microlensing-induced flux variability is a robust way to test whether Godzilla could be a single star or a system of a few stars \citep{Choe2024}.

Taking into account both flux ratios and flux variability limits, we conclude that the scenario of a young star cluster magnified by somewhere around one thousand fold is viable and is compatible with the observed lensing properties of Godzilla. If Godzilla is a young star cluster, its magnification should be less than about $\mu=2000$ and its stellar mass higher than $1 \times 10^6\,M_\odot$.


\section{Conclusions}
\label{sec:concl}

High lensing magnifications have offered a unique view of the impressive star formation activities on parsec scales in the Cosmic Noon galaxy Sunburst. Apart from the 12-imaged Lyman-continuum-leaking super star cluster that has been previously studied, in this work we study Godzilla, which is seemingly a more perplexing source. We suggest that Godzilla is a young super star cluster with its core enshrouded with dense massive star ejecta photo-excited by stellar ionizing radiation. We have combined HST filter photometry and nebular emission line measurements from VLT/MUSE and X-shooter spectroscopy to model the star cluster and the associated nebula. 

We find a cluster age $4$--$6\,$Myr and a likely stellar metallicity $Z=0.25\,Z_\odot$ for Godzilla. For a standard stellar IMF and depending on the total and tangential magnification factors $\mu$ and $\mu_t$, we find a large cluster stellar mass $M_\star = 2 \times 10^6\,M_\odot\,(1000/\mu)$, and a PSF-based size constraint for its FUV-bright component, $R_{\rm FUV}\lesssim 1\,{\rm ~pc}\, (500/\mu_t)$. The high compactness should place Godzilla in a regime where rapid radiative cooling renders stellar wind and CCSNe material trapped in the cluster potential regardless of the initial ejecta speed. Indeed, despite the inferred SMC-like stellar metallicity, we found evidence for gas-phase He (He/H$=0.27^{+0.09}_{-0.10}$) and N ($\log{\rm N/O}=-0.31^{+0.08}_{-0.08}$) enrichment indicative of massive star winds, as well as gas-phase O enhancement ($12+\log{\rm O/H}=8.75^{+0.19}_{-0.25}$) indicative of CCSNe not long after onset. We also find sub-solar C/O, Ne/O and Si/O values in the nebula, which may have implications for CCSN light element yields for progenitor ZAMS masses greater than $40\,M_\odot$.

Through a comparison to models of star cluster ejecta from \citetalias{Molla2012StarClusterEjecta}, we confirm that the inferred gas-phase abundances are consistent with complete retention of stellar winds and CCSN ejecta up to a cluster age around $4$--$6\,$Myr (\reffig{molla} and \reffig{molla_o}), which also implies that stars more massive than $m_{\rm ZAMS} = 40\,M_\odot$ have successfully exploded and enriched the intracluster medium. We note that our inferred $\log{\rm Ne/O}=-1.13^{+0.16}_{-0.13}$ is lower than the model prediction of \citetalias{Molla2012StarClusterEjecta} and is interestingly lower than the solar value. While a large sample of less extreme star-forming regions show a solar Ne/O value independent of metallicity, Godzilla appears young enough that only stars more massive than about $40\,M_\odot$ have reached the end of their lives. We therefore posit that the low Ne/O may be related to the CCSN yields of such massive progenitors, with speculations on photodisintegration of Ne to O through explosive nucleosynthesis and collapse of Ne-Mg-O shell into the remnant.

The nebula of Godzilla is remarkably dense $\nele \sim 10^{7–8}\,{\rm cm}^{-3}$, which indicates an extremely high intracluster pressure $P \sim 10^{11–12}\,{\rm K}\,{\rm cm}^{-3}$. We suggest that the high pressure is provided by supersonic turbulence in stellar ejecta that is accumulated in the cluster's gravitational potential due to runaway radiative cooling, which is perhaps only a fraction of a parsec across (\refsec{geometry}). The retained stellar ejecta is dense enough to self-shield against ionizing photons, is likely dust-free, and stays warm by UV heating. Through simple analytic analysis of CCSN kinetic feedback within a dense medium, we find that CCSNe after an age $\sim 3\,$Myr, is possibly capable of maintaining this turbulence, but probably have failed to evacuate this enshrouding dense gas out of the cluster (\refsec{driving}). We theorize that nebular emission lines originate from many ionization-bounded hot bubbles that are blown either by massive star winds or by past CCSNe (see \reffig{cartoon}). We have argued that such unusual geometry with a significant HI column along the line of sight may explain the weakness of H Balmer lines (\refsec{yeffect}) and the observed Ly$\alpha$-pumped Fe fluorescent emission (\refsec{bowen}). 

Recent JWST NIRCam and NIRSpec data (PID: 2555, PI: Rivera-Thorsen) revealed optical emission lines with both narrow and broad components, many of which had only conservative upper limits in this work. The UV emission lines show no evidence of such a broad component, and we speculate a lower density, outflowing gas component which dominates the {\rm [O III]}$\lambda\lambda$4959,5007 and H Balmer lines may be a natural explanation for this emission in the context of our physical model. Further analysis of these data will test this scenario (Pascale et al., in preparation), and upcoming JWST MIRI imaging (PID: 6353, PI: Pascale) will allow for the best panchromatic view of this exciting object to date.

We have reminded the readers of the difficulty in accurately determining the magnification factor $\mu$ for Godzilla through galaxy-cluster scale lens modeling, in which it is challenging to include optically undetectable substructure lenses. We have discussed the possible value of $\mu$ from two considerations. For one, a lower limit on the magnification ratio between Godzilla and a possibly blended counter lensed image leads us to conclude that $\mu \gtrsim 500$ for Godzilla. Another consideration is based on flux variability expected from intracluster microlensing effects acting on member stars. We have derived stringent limits on the flux variability using archival HST imaging data over 18 observed months, which implies $\mu < 2000$ if Godzilla is a young star cluster. We have made the $\mu$ dependence explicit in those of our quantitative results that are sensitive to the value of $\mu$, and have presented numerical results corresponding to a fiducial $\mu=1000$. Our physical picture of condensed gas in a young star cluster is qualitatively robust for a wide range of magnification values $300 < \mu < 2000$. We further note that the stringent limits on flux variability are in tension with the alternative scenario that Godzilla is a magnified single star with $\mu > 7000$, unless the source's effective velocity relative to the caustic is one order of magnitude smaller than the most probable value.

Godzilla may be an analog of globular cluster progenitor at SMC-like metallicity. It is interesting to ponder whether the observed retained cluster gas would birth a 2P stellar population. However, the enhanced gas-phase O/H is not consistent with the typical abundance pattern of 2P stars. The gas-phase He/H elevation is consistent with pollution by evolved star winds, but is likely excessive compared to the typical He abundance variations seen in GCs. Altogether this implies that, should Godzilla be a true GC progenitor and host multiple stellar populations, the 2P stars probably have already formed, prior to the onset of evolved star winds and CCSNe which would dump excessive He and O respectively. This appears to align with very massive main sequence stars or supermassive stars as the polluters for multiple stellar population formation. Finding more objects similar to Godzilla will further shed light on these intriguing questions.

\section*{acknowledgments}
Some of the data used in this paper were obtained from the Mikulski Archive for Space Telescopes (MAST) at the Space Telescope Science Institute. The specific observations analyzed can be accessed via this \dataset[DOI]{https://doi.org/10.17909/tznp-gv14}.

We are foremost grateful to Christopher McKee for the countless inspiring discussions we had with him throughout the course of this work. We would like to also thank Luca Boccioli, Sanjana Curtis, Neal Dalal, Michael Fall, Xiao Fang, Brenda Frye, Mike Grudi\'{c}, Alexander Ji, Chiaki Kobayashi, Wenbin Lu, Raffaella Margutti, Christopher Matzner, Shyam Menon, and Benny Tsang for the very valuable discussions and comments. We also thank the anonymous referee whose valuable feedback greatly helped us improve the quality of this work.

MP acknowledges funding support through the NSF Graduate Research Fellowship grant No.~DGE 1752814, and acknowledges the support of System76 for providing computer equipment. L.D. acknowledges research grant support from the Alfred P. Sloan Foundation (Award Number FG-2021-16495), and support of Frank and Karen Dabby STEM Fund in the Society of Hellman Fellows.



\bibliography{godzilla}{}
\bibliographystyle{aasjournal}



\end{document}